\documentclass[twocolumn,showpacs,amsfonts,aps,prc,floatfix,%
nofootinbib,superscriptaddress]{revtex4} 
\usepackage{amsmath}
\usepackage{bm}
\usepackage{graphicx}

\newcommand{\Tdec}{T_\mathrm{dec}}
\newcommand{\edec}{e_\mathrm{dec}}
\newcommand{\ecc}{\varepsilon}
\newcommand{\Tc}{T_\mathrm{c}}
\newcommand{\dNdy}{dN_\mathrm{ch}/dy}

\newcommand{\be}[1]{\begin{equation}\label{#1}}
\newcommand{\ee}{\end{equation}}

\newcommand{\eq}{{\,=\,}}

\def\La{\langle}
\def\Ra{\rangle}

\usepackage{color}

\begin{document}

%%%%%%%%%%%%%%%%%%%%%%%%Front Matter%%%%%%%%%%%%%%%%%%%%%%%%%%%%%%%%%%
%%%%%%%%%%%%%%%%%%%%%%%%%%%%%%%%%%%%%%%%%%%%%%%%%%%%%%%%%%%%%%%%%%%%%%

\title{Multiplicity scaling in ideal and viscous hydrodynamics}
\date{\today}

\author{Huichao Song}
\email[Correspond to\ ]{song@mps.ohio-state.edu}
\affiliation{Department of Physics, The Ohio State University, 
%191 West Woodruff Avenue, 
Columbus, OH 43210, USA}
\author{Ulrich Heinz}
\affiliation{Department of Physics, The Ohio State University, 
%191 West Woodruff Avenue, 
Columbus, OH 43210, USA}
\affiliation{CERN, Physics Department, Theory Division, CH-1211 Geneva 23, 
             Switzerland}

\begin{abstract}
Using numerical results from ideal and viscous relativistic hydrodynamic
simulations with three different equations of state, for Au+Au and 
Cu+Cu collisions at different centralities and initial energy densities, 
we explore the dependence of the eccentricity-scaled elliptic flow, 
$v_2/\varepsilon$, and the produced entropy fraction, $\Delta {\cal S}/
{\cal S}_0$, on the final charged hadron multiplicity density $\dNdy$ per 
unit transverse overlap area $S$, $(1/S)\dNdy$. The viscous 
hydrodynamic simulations are performed with two different versions of 
the Israel-Stewart kinetic evolution equations, and in each case 
we investigate the dependence of the physical observables on the 
kinetic relaxation time. We find approximate scaling of 
$v_2/\varepsilon$ and $\Delta S/S_0$ with $(1/S)\dNdy$, with scaling 
functions that depend on the EOS and, in particular, on the value of the 
specific shear viscosity $\eta/s$. Small scaling violations are seen 
even in ideal hydrodynamics, caused by a breaking of the scale invariance 
of ideal fluid dynamics by the freeze-out condition. Viscous 
hydrodynamics shows somewhat larger scale-breaking effects that increase 
with increasing $\eta/s$ and decreasing system size and initial energy 
density. We propose to use precision studies of these scaling violations 
to help constrain the shear viscosity $\eta/s$ of the quark-gluon plasma 
created in relativistic heavy ion collisions.
\end{abstract}
\pacs{25.75.-q, 12.38.Mh, 25.75.Ld, 24.10.Nz}

\maketitle

%%%%%%%%%%%%%%%%%%%%%%%%%%%%%%%%%%%%%%%%%%%%%%%%%%%%%%%%%%%%%%%%%%%%%%%%%%%%
\section{Introduction}\label{sec1}
%%%%%%%%%%%%%%%%%%%%%%%%%%%%%%%%%%%%%%%%%%%%%%%%%%%%%%%%%%%%%%%%%%%%%%%%%%%%

Recent numerical studies of relativistic hydrodynamics for dissipative
fluids \cite{Romatschke:2007mq,Chaudhuri:2007zm,Song:2007fn,%
Dusling:2007gi} have confirmed earlier estimates \cite{Heinz:2002rs,%
Teaney:2003kp} that the ``elliptic'' anisotropic collective flow 
observed in non-central heavy-ion collisions is very sensitive to the 
shear viscosity of the matter formed in such collisions. Since ideal 
fluid dynamics (i.e. the assumption that viscosity can be neglected) 
provides a phenomenologically quite successful description of much of 
the soft hadron data collected from Au+Au collisions at the 
Relativistic Heavy Ion Collider (RHIC) \cite{experiments,reviews,%
Heinz:2004ar}, this implies strong constraints on the shear viscosity to 
entropy ratio $\eta/s$ \cite{Teaney:2003kp,Lacey:2006bc} and the 
thermalization time scale \cite{Heinz:2001xi,Heinz:2004pj} of the matter 
in the collision fireball. The conclusion is that the quark-gluon plasma 
(QGP) created at RHIC is a strongly coupled plasma with almost perfect 
liquid collective behavior \cite{Gyulassy:2004vg,Gyulassy:2004zy,%
Shuryak:2004cy} whose specific shear viscosity is lower than that of 
any previously known real fluid and consistent with a postulated lower 
bound of $\eta/s\geq\hbar/(4\pi k_B)$ derived from the study of 
infinitely strongly coupled conformal field theories \cite{Ads-CFT,son-PRL05}
using the AdS/CFT correspondence and corroborated by earlier quantum 
mechanical arguments based on the uncertainty relation \cite{Gyulassy85}.

On the other hand, heavy-ion data at RHIC, SPS and AGS also show that 
ideal hydrodynamics gradually breaks down at larger impact parameters,
for smaller collision systems, at lower collision energies, and away 
from midrapidity (see \cite{Heinz:2004ar} for a review). Much of this 
can be attributed to strong viscous effects in the late collision stage 
after the QGP has converted to hadrons \cite{Hirano:2005xf}. However, 
due to uncertainties in the initial conditions for the fireball 
deformation, there is some room left for a non-zero value of the QGP 
viscosity \cite{Hirano:2005xf}. To study this further requires a 
viscous hydrodynamic approach because the tool used to describe the 
non-equilibrium late hadronic dynamics (a classical cascade) is not
well suited for the rapidly evolving, very dense matter in the early 
collision stages

The motivation for the present paper is provided by the well-known 
systematic comparison of Voloshin {\it et al.} \cite{Alt:2003ab,VoloshinQM06}
of elliptic flow data with ideal fluid dynamical predictions which 
suggests that the elliptic flow parameter $v_2$ scaled by the initial
source eccentricity $\varepsilon$, $v_2/\ecc$, while strongly deviating
from ideal hydrodynamics at low multiplicities, still scales with 
the final multiplicity per unit overlap area:
\begin{eqnarray}
\label{v2}
  \frac{v_2}{\ecc} \propto \frac{1}{S}\frac{dN_\mathrm{ch}}{dy}.
\end{eqnarray}
For ideal fluids the right hand side is a direct measure of the initial 
entropy density \cite{Hwa:1985xg}. The scaling (\ref{v2}) implies that all 
dependence on impact parameter, collision energy and system size can be, 
to good approximation, absorbed by simply taking into account how these 
control parameters change the final hadron multiplicity density. We will 
call this observation simply ``multiplicity scaling of the elliptic flow'',
where ``elliptic flow'' is a shorthand for the eccentricity-scaled 
elliptic flow $v_2/\ecc$ and ``multiplicity'' stands for 
$(1/S)\dNdy$. 

Such a scaling is expected for ideal fluid dynamics whose equations of 
motion are scale invariant and where the eccentricity-scaled elliptic
flow is therefore predicted \cite{Ollitrault:1992bk,Bhalerao:2005mm}
to depend only on the squared speed of sound, $c_s^2=\frac{\partial p}
{\partial e}$, 
which describes the stiffness of the equation of state (EOS) or ``pushing 
power'' of the hydrodynamically expanding matter. It has been known,
however, for many years \cite{Kolb:1999it} that this ideal-fluid scaling is 
broken by the final freeze-out of the matter: if hadron freeze-out is 
controlled by hadronic cross sections (mean free paths) or simply
parametrized by a critical decoupling energy density $\edec$
or temperature $\Tdec$, this introduces and additional scale
into the problem which is independent of (or at least not directly 
related to) the initial geometry of the fireball. This breaks the 
above argument based on scale invariance of the ideal fluid equations of
motion. We will show here that this also leads to a breaking of the 
multiplicity scaling of $v_2/\ecc$ not only in the most peripheral 
or lowest energy collisions, where freeze-out obviously cuts the 
hydrodynamic evolution short since the freeze-out density is reached 
before the flow anisotropy can fully build up \cite{Kolb:1999it}, but 
even in the most central collisions at RHIC where freeze-out still 
terminates the hydrodynamic evolution before the elliptic flow can 
fully saturate (see also \cite{Hirano:2005xf}).

The more interesting aspect of the experimentally observed scaling is,
however, its apparent validity in regions where ideal fluid dynamics
does not work (these encompass most of the available data \cite{Alt:2003ab}).
Many years ago, simple scaling laws for the centrality dependence
of elliptic flow were derived from kinetic theory in the dilute gas 
limit, where the particles in the medium suffer at most one 
rescattering before decoupling \cite{Heiselberg:1998es,Voloshin:1999gs}; 
these can be reinterpreted in terms of multiplicity scaling for 
$v_2/\ecc$. The dilute gas limit is expected to hold for very small 
collision systems, very large impact parameters or very low collision 
energies. More recently, a successful attempt was made to 
phenomenologically connect the dilute gas and hydrodynamic limits
with a 1-parameter fit involving the Knudsen number \cite{Drescher:2007cd}.
This fit works very well for Au+Au and Cu+Cu data from RHIC, but
predicts that even in the most central Au+Au collisions at RHIC the
ideal fluid dynamical limit has not yet been reached and is missed by
at least 25\% \cite{Drescher:2007cd}. In the present paper we use
viscous relativistic hydrodynamics to explore the multiplicity scaling
of $v_2/\ecc$ in the phenomenologically relevant region. We conclude
(not surprisingly since much of the available data is from regions
where the viscous hadronic phase plays a large role \cite{Hirano:2005xf}) 
that the multiplicity scaling data \cite{Alt:2003ab,VoloshinQM06} require 
significant shear viscosity for the medium, but also that viscous 
hydrodynamics predicts subtle scaling violations which seem to be 
qualitatively consistent with trends seen in the data (even if the 
experimental evidence for scaling violations is presently not statistically 
robust) and whose magnitude is sensitive to the specific shear viscosity 
$\eta/s$. This gives hope that future more precise data can help
constrain the QGP shear viscosity through exactly such scaling 
violations. 

We should caution the reader that, similar to Ref. \cite{Drescher:2007cd}
which used a constant (time-independent) Knudsen number, our viscous
hydrodynamic calculations are done with a constant (temperature-independent)
specific shear viscosity $\eta/s$. Neither assumption is realistic, and 
we expect $\eta/s$ in particular to show strong temperature dependence
near $\Tc$ (the critical temperature for the quark-hadron phase
transition) and emerge from the phase transition with much larger values
than in the QGP phase. Comparisons between the results presented here
and experimental data are therefore, at best, indicative of qualitative
trends, and improved calculations, which in particular match viscous 
hydrodynamics to a realistic hadron cascade below $\Tc$, are required 
before an extraction of $\eta/s$ from experimental data can be attempted.  

%%%%%%%%%%%%%%%%%%%%%%%%%%%%%%%%%%%%%%%%%%%%%%%%%%%%%%%%%%%%%%%%%%%%%%%%%%%
\section{Dissipative fluid dynamics}\label{sec2}
%%%%%%%%%%%%%%%%%%%%%%%%%%%%%%%%%%%%%%%%%%%%%%%%%%%%%%%%%%%%%%%%%%%%%%%%%%%

In this section we briefly review the viscous hydrodynamic equations 
that we solve, focussing on some differences in the formulations used in
previously published papers \cite{Romatschke:2007mq,Song:2007fn,%
Dusling:2007gi,Chaudhuri:2007zm} which we investigate here 
further. For technical details we ask the reader to consult these 
earlier papers.

We focus on systems with exact longitudinal boost-invariance and use
the code {\tt VISH2+1} \cite{Song:2007fn} to solve numerically for the 
expansion in the two dimensions transverse to the beam direction. 
As in \cite{Romatschke:2007mq,Song:2007fn,Dusling:2007gi,%
Chaudhuri:2007zm} we consider only shear viscosity, neglecting bulk 
viscosity and heat conduction. (Bulk viscosity may become large near
the QCD phase transition \cite{Paech:2006st,Kharzeev:2007wb,Meyer:2007dy}
and should thus be included in the future before comparing viscous 
hydrodynamics with experimental data.) {\tt VISH2+1} solves the 
conservation laws for energy and momentum, $d_m T^{mn}\eq0$ (where 
$d_m$ is the covariant derivative in our curvilinear $(\tau,x,y,\eta)$ 
coordinate system \cite{Heinz:2005bw}), with the decomposition
\begin{eqnarray}
 T^{mn} = e u^m u^n - p\Delta^{mn} + \pi^{mn},\
 \Delta^{mn} =g^{mn}{-}u^m u^n,
\end{eqnarray}
together with evolution equations for the viscous shear pressure tensor 
components $\pi^{mn}$: 
\begin{eqnarray}
\label{EQpimn}
 D\pi^{mn} &=&-\frac{1}{\tau_{\pi}}(\pi^{mn}{-}2\eta\sigma^{mn})
   -(u^m\pi^{nk}{+}u^n\pi^{mk}) Du_k
\nonumber\\
   && -\frac{1}{2}\pi^{mn} \frac{\eta T}{\tau_\pi}
       d_k\left(\frac{\tau_\pi}{\eta T}u^k\right)
    - \pi_a^{\,(m}\omega^{n)a}.
\end{eqnarray}
Here $D{\eq}u^m d_m$ is the time derivative in the local comoving
frame, $\nabla^m\eq\Delta^{ml}d_{l}$ is the spatial gradient in that
frame, and $\sigma^{mn}\eq\nabla^{\left\langle m\right.} u^{\left.n
\right\rangle}\eq\frac{1}{2}(\nabla^m u^n{+}\nabla^n u^m)-\frac{1}{3}
\Delta^{mn}\theta$ (with the scalar expansion rate $\theta\equiv d_k u^k
=\nabla_k u^k$) is the symmetric and traceless velocity shear tensor.
$\omega_{mn}=\nabla_n u_m-\nabla_m u_n$ is the vorticity tensor, and 
$a^{(m}b^{n)}\equiv$ $\frac{1}{2}(a^m b^n+a^n b^m)$ denotes symmetrization.
Even though several components of the symmetric shear pressure tensor 
$\pi^{mn}$ are redundant \cite{Heinz:2005bw} on account of its 
tracelessness and transversality to the flow velocity $u^m$, {\tt VISH2+1} 
propagates all 7 non-zero components and uses the tracelessness and
transversality conditions as checks of the numerical accuracy
\cite{Song:2007fn}.

For a conformally symmetric fluid such as a massless quark-gluon gas, 
the temperature $T$ is the only scale in the problem and therefore 
$\eta\sim s\sim T^3$ and $\tau_\pi\sim 1/T$, hence $\eta T/\tau_\pi 
\sim T^5$. In this limit the first term in the second line of 
Eq.~(\ref{EQpimn}) can be written as \cite{Muronga:2004sf}
\begin{equation}
\label{confid}
  -\frac{1}{2}\pi^{mn} \frac{\eta T}{\tau_\pi}
         d_k\left(\frac{\tau_\pi}{\eta T}u^k\right)
  = +\frac{1}{2}\pi^{mn}\bigl(5 D(\ln T)-\theta\bigr)\,.
\end{equation}
This is the form used in Ref.~\cite{Romatschke:2007mq}. 

It has recently been argued \cite{Baier:2007ix,Bhattacharyya:2008jc,%
Natsuume:2007ty,Loganayagam:2008is} that the r.h.s. of Eq.~(\ref{EQpimn}) 
should contain even more terms, at least for conformal fluids in the 
strong coupling limit. We will not pursue this possibility here.

Equations (\ref{EQpimn}) are known as ``Israel-Stewart (I-S) equations''
and based on an expansion of the entropy production rate to second order 
\cite{Israel:1976tn,Muronga:2001zk,Muronga:2003ta,Muronga:2004sf,%
Muronga:2006zw} (macroscopic approach) or, in a microscopic kinetic 
approach using Grad's 14-moment method, of the phase-space distribution 
function to first order in small deviations from local thermal equilibrium 
\cite{Israel:1976tn,Muronga:2006zw,Baier:2006um} (see also \cite{Muller}). 
By introducing a finite and sufficiently large relaxation time $\tau_\pi$ 
for the evolution of the shear pressure tensor towards its Navier-Stokes 
limit $\pi^{mn}=2\eta\sigma^{mn}$, these equations eliminate problems 
with acausal signal propagation at short wavelengths and the resulting 
numerical instabilities that famously plague the relativistic 
Navier-Stokes equation. A somewhat different approach to solving these 
problems was developed by \"Ottinger and Grmela (\"O-G) \cite{OG} and 
has been used in \cite{Dusling:2007gi}; because a comparison of results 
obtained with the I-S and \"O-G equations is non-trivial, we will leave 
that for a later study. 

Refs.~\cite{Romatschke:2007mq,Song:2007fn,Chaudhuri:2007zm}
use different versions of Eqs.~(\ref{EQpimn}). P. \& U. Romatschke 
\cite{Romatschke:2007mq} use the full set of terms displayed in 
(\ref{EQpimn}) which we label as ``full I-S equation''. The last term 
in the second line of Eq.~(\ref{EQpimn}) involving the vorticity cannot 
be obtained from the macroscopic approach \cite{Muronga:2003ta,Heinz:2005bw} 
since it does not contribute to entropy production, but it follows from 
the microscopic kinetic approach \cite{Israel:1976tn,Baier:2006um}. For 
longitudinally boost-invariant systems with initially vanishing transverse 
flow it is zero initially and was found in \cite{Romatschke:2007mq} to 
remain tiny throughout the fireball evolution. We can therefore remove it 
from consideration when comparing published results from the different 
approaches. The first term in the second line of (\ref{EQpimn}) 
arises in this form from the macroscopic approach (2$^\mathrm{nd}$ order
entropy production \cite{Muronga:2003ta,Heinz:2005bw}) but was
neglected in our previous work \cite{Song:2007fn}, following
an argument in \cite{Heinz:2005bw} that it is of second order in small
quantities and therefore subdominant compared to the first two terms
on the r.h.s. of Eq.~(\ref{EQpimn}). A similar argument can be made for 
the last term in line 1 of (\ref{EQpimn}) (which does not contribute to
entropy production either), but Baier {\it et al.} \cite{Baier:2006um} 
pointed out that this term is needed to preserve the transversality of 
$\pi^{mn}$ during kinetic evolution. In Ref.~\cite{Song:2007fn} 
we therefore kept all terms in the {\em first} line of Eq.~(\ref{EQpimn})
but dropped those in the {\em second} line; we call this here the 
``simplified I-S equation''. In Ref.~\cite{Baier:2007ix} Baier {\it et al.} 
argued that for a conformally invariant medium, such as a classical 
massless quark-gluon gas, the first term in the {\em second} line is 
needed to preserve the conformal invariance of the kinetic evolution 
equation and hence should not be dropped. We now understand that the 
arguments presented in \cite{Heinz:2005bw} to neglect all but the first 
term on the right hand side of Eq.~(\ref{EQpimn}) were at best 
superficial since this term involves the difference between two 
first-order quantities and thus presumably needs to be counted as 
small of second order.

Chaudhuri followed in his work \cite{Chaudhuri:2007zm} the approach 
advertised in \cite{Heinz:2005bw}; as a result, in his procedure 
$\pi^{mn}$ must be expected to evolve away from transversality. He
circumvents this problem by evolving only the three linearly 
independent components of $\pi^{mn}$ and computing the rest from the
tracelessness and transversality conditions \cite{Chaudhuri:2007zm}. 
The problem resurfaces, however, since now the results for all components
of $\pi^{mn}$ must be expected to depend on the choice of independent 
components which are evolved dynamically with the truncated equation 
(\ref{EQpimn}). While these questions await quantitative study we note 
that Chaudhuri's results \cite{Chaudhuri:2007zm} appear to differ 
significantly from ours \cite{Song:2007fn}.

In the present paper we show many comparisons between solutions obtained
by using the ``full I-S equations'' with those from the ``simplified 
I-S equations''. Although at sufficiently long wavelengths both have to 
agree in the Navier-Stokes limit $\tau_\pi\to0$ (up to issues of 
numerical stability), as inspection of Eq.~(\ref{EQpimn}) readily shows,
they differ for non-zero $\tau_\pi$ and will be seen to exhibit different 
degrees of sensitivity to $\tau_\pi$. This is of phenomenological importance
since $\tau_\pi$ for the QGP is not known, and a strong sensitivity to this
unknown parameter will negatively impact our ability to extract the QGP
shear viscosity $\eta/s$ from experimental data.
 
%%%%%%%%%%%%%%%%%%%%%%%%%%%%%%%%%%%%%%%%%%%%%%%%%%%%%%%%%%%%%%%%%%%%%%%%%%%
\section{Initial conditions, freeze-out, and equation of state}\label{sec3}
%%%%%%%%%%%%%%%%%%%%%%%%%%%%%%%%%%%%%%%%%%%%%%%%%%%%%%%%%%%%%%%%%%%%%%%%%%%

For the present study, we initialize the expanding fireball in the same 
way as in Ref.~\cite{Song:2007fn}, i.e. with vanishing initial transverse 
flow and with an initial energy density profile proportional to the 
transverse density of wounded nucleons, calculated from a Saxon-Woods 
nuclear density profile with radius and surface thickness parameters 
$R_0=4.2$\,fm, $\xi=0.596$\,fm for Cu and $R_0=6.37$\,fm, $\xi=0.56$\,fm 
for Au nuclei. The energy density profile is normalized by a parameter 
$e_0=e(\tau_0;r{=}b{=}0)$ giving the peak energy density in the center 
of the fireball for central collisions (impact parameter $b=0$). $e_0$ 
is related to the peak wounded nucleon density in the same collisions 
by a factor $\kappa$ which is assumed to depend on energy but not on 
the size of the colliding nuclei. We here consider $e_0$ values that
lead to final multiplicities covering the range accessible at RHIC
and beyond, albeit perhaps not all the way to the Large Hadron Collider
(LHC). 

As of now, the energy dependence of $\kappa$ cannot be calculated and must 
be determined empirically from the final charged hadron multiplicity
$\dNdy$. Since $\dNdy$ counts the final entropy per unit of rapidity, 
including any entropy generated by viscous effects during the expansion, 
the value of $\dNdy$ corresponding to a given $\kappa$ will depend on 
the viscosity $\eta/s$. We will see that the amount of entropy produced
by viscous effects additionally depends on system size, impact parameter 
and collision energy, but that all these dependences can, to good 
approximation, be absorbed in a single scaling function, with parametric
dependence on $\eta/s$, that depends only on the multiplicity density
$(1/S)\dNdy$: similar to $v_2/\ecc$, entropy production 
$\Delta{\cal S}/{\cal S}_0$ exhibits approximate ``multiplicity scaling''. 
However, this scaling function turns out to be non-linear. It therefore 
modifies the centrality dependence of the produced charged multiplicity, 
softening the observed increase with collision centrality of the produced 
charged multiplicity per pair of wounded nucleons, $2\frac{\dNdy}
{N_\mathrm{part}}$. Exploration of this important issue requires an 
accurate modeling of 
the impact parameter dependence of the initial entropy density profile 
using, say, the Glauber or color glass condensate models. This is 
beyond the scope of the present article and will be left for a future 
study.

Following the majority of previous studies \cite{Chaudhuri:2007zm,%
Song:2007fn,Dusling:2007gi,Chaudhuri:2005ea,Baier:2006gy}, the viscous 
shear pressure tensor is initialized with its Navier-Stokes value 
$\pi^{mn}=2\eta\sigma^{mn}$. Other initial conditions were studied
in \cite{Romatschke:2007mq,Song:2007fn}, but the final observables 
were found to be insensitive to such variations \cite{Song:2007fn}.  
The kinetic relaxation time $\tau_\pi$ for the kinetic evolution of 
the shear pressure tensor is taken as $\tau_\pi=
c_\pi\tau_\pi^\mathrm{Boltz}$ where $\tau_\pi^\mathrm{Boltz} = 
\frac{6}{T}\frac{\eta}{s}$ is the kinetic theory value for a classical
gas of massless Boltzmann particles \cite{Israel:1976tn} and $c_\pi$
is varied between $\frac{1}{4}$ and 1.

Decoupling from the hydrodynamic fluid is implemented by following the 
same procedure as described in \cite{Song:2007fn}. We use the {\tt AZHYDRO} 
algorithm \cite{AZHYDRO} to find the freeze-out surface at constant
decoupling temperature $\Tdec=130$\,MeV and calculate the final
hadron spectra from the Cooper-Frye integral over this surface 
\cite{Cooper-Frye}, with a distribution function that accounts for
the remaining small deviations from local thermal equilibrium along
that surface \cite{Teaney:2003kp,Song:2007fn}. Resonance decays are
neglected, and only the elliptic flow of directly emitted pions is 
shown. To estimate the total charged hadron multiplicity, we take the 
directly emitted positive pions, multiply by 1.5 to roughly account for 
multiplication by resonance decays at $\Tdec$, then multiply by another 
factor $2\times1.2=2.4$ to account for the negatives and roughly 20\% 
of final charged hadrons that are not pions. A proper calculation of 
the resonance decay chain is computationally expensive and, for a 
systematic study like the one presented here that requires hundreds of 
runs of {\tt VISH2+1}, beyond our presently available resources. 

%
%%%%%%%%%%%%%%%%%%%%%%%%%%%%%% Fig. 1 EOS %%%%%%%%%%%%%%%%%%%%%%%%%%%%%%%%%%
\begin{figure}[htb]
\includegraphics[width=\linewidth,clip=]{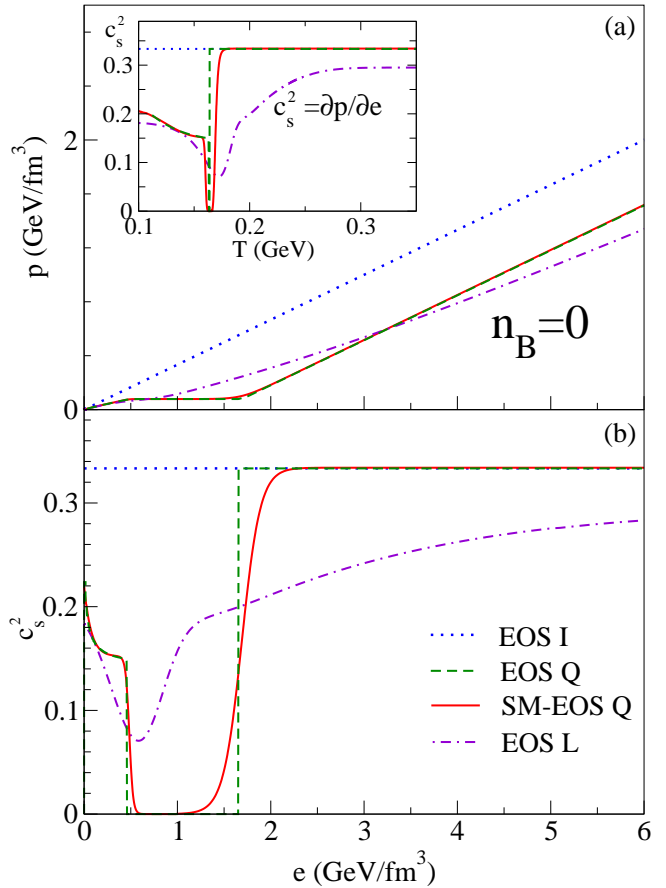}\\
\caption{(Color online) The equation of state. Panel (a) shows the
pressure $p$ as a function of energy density $e$ and (in the inset) the 
squared speed of sound $c_s^2=\frac{\partial p}{\partial e}$ as a function
of temperature $T$. Panel (b) shows $c_s^2$ as a function of energy 
density $e$.} 
\label{F1}
\end{figure}
%%%%%%%%%%%%%%%%%%%%%%%%%%%%%%%%%%%%%%%%%%%%%%%%%%%%%%%%%%%%%%%%%%%%%%%%%%%%
%

Figure~\ref{F1} shows the three equations of state (EOS) explored
in the present study. EOS~I describes a non-interacting gas of massless
particles, $e=p/3$. EOS~Q is a frequently employed equation of state 
\cite{Kolb:1999it} that matches a non-interacting quark-gluon gas above 
$T_c$ in a first-order transition (Maxwell construction) to a realistic 
hadron resonance gas (HRG) in chemical equilibrium below $T_c$, using a 
bag parameter $B$ to adjust $T_c$ to $T_c=164$\,MeV. SM-EOS~Q is a 
slightly smoothed version of EOS~Q, see \cite{Song:2007fn} for details. 
Since the discontinuities in the function $c_s^2(e)$ for EOS~Q cause 
numerical problems in {\tt VISH2+1} due to large velocity gradients near \
the interfaces between QGP, mixed phase and HRG, we here use SM-EOS~Q. 
 
EOS~L matches the same hadron resonance gas below $T_c$ smoothly in a 
rapid cross-over transition to lattice QCD data \cite{Katz05} above 
$T_c$. For the fit, the lattice data were plotted in the form $p(e)$, 
interpolated and then smoothly joined to the $p(e)$ curve of the HRG. 
As can be seen in the upper panel of Fig.~\ref{F1} in the inset, our 
procedure is not fully thermodynamically consistent and leads to a 
somewhat different temperature dependence of $c_s^2$ below $T_c$ than 
for EOS~Q and SM-EOS~Q. Since this only affects the flow dynamics below 
our decoupling temperature of $\Tdec=130$\,MeV, we have not put any 
effort into correcting this. For future comparisons of viscous 
hydrodynamic calculations with experimental data, the chemical equilibrium 
hadron resonance gas below $T_c$ employed here must be replaced by a
chemically non-equilibrated hadron gas whose particle ratios are frozen
in at the chemical decoupling temperature $T_\mathrm{chem}\approx T_c$;
this has well-known consequences for the final hadron spectra and elliptic
flow which can not be neglected \cite{Hirano:2005wx}. We postpone this,
together with a more careful and thermodynamically fully consistent 
matching to the lattice QCD data, to a future study. We note, however,
that EOS~L shown in Fig.~\ref{F1} is quite similar to ``EOS~qp'' studied in 
Ref.~\cite{Huovinen:2005gy}.

%%%%%%%%%%%%%%%%%%%%%%%%%%%%%%%%%%%%%%%%%%%%%%%%%%%%%%%%%%%%%%%%%%%%%%%%%%%%%%
\section{Evolution of momentum anisotropies: simplified vs. full I-S
equations}
\label{sec4}
%%%%%%%%%%%%%%%%%%%%%%%%%%%%%%%%%%%%%%%%%%%%%%%%%%%%%%%%%%%%%%%%%%%%%%%%%%%%%%

It has been previously observed that the results of 
Refs.~\cite{Romatschke:2007mq} and \cite{Song:2007fn} for the differential
elliptic flow $v_2(p_T)$, although both based on the Israel-Stewart 
2$^\mathrm{nd}$ order formalism, seemed to disagree, our work 
\cite{Song:2007fn} showing much stronger viscous suppression of $v_2$ 
than that of P. \& U.~Romatschke \cite{Romatschke:2007mq}. The resolution 
of this discrepancy was made difficult by the fact that the two groups 
not only used different versions of the Israel-Stewart equation 
(\ref{EQpimn}) as described in Sec.~\ref{sec2}, but also different 
initial conditions, different equations of state, and system sizes 
(Cu+Cu \cite{Song:2007fn} vs. Au+Au \cite{Romatschke:2007mq}). In 
\cite{Song:2007fn} we noted in a footnote that the main reason for 
the observed differences seemed to be the different I-S equations used 
by the two groups. As we will see, this is only part of the story. In 
this sections we explore this question further and lead it to a complete 
resolution. 

%
%%%%%%%%%%%%%%%%%%%%%%%%%%%%%% Fig. 2 \ecc_p %%%%%%%%%%%%%%%%%%%%%%%%%%%%%%%%
\begin{figure*}
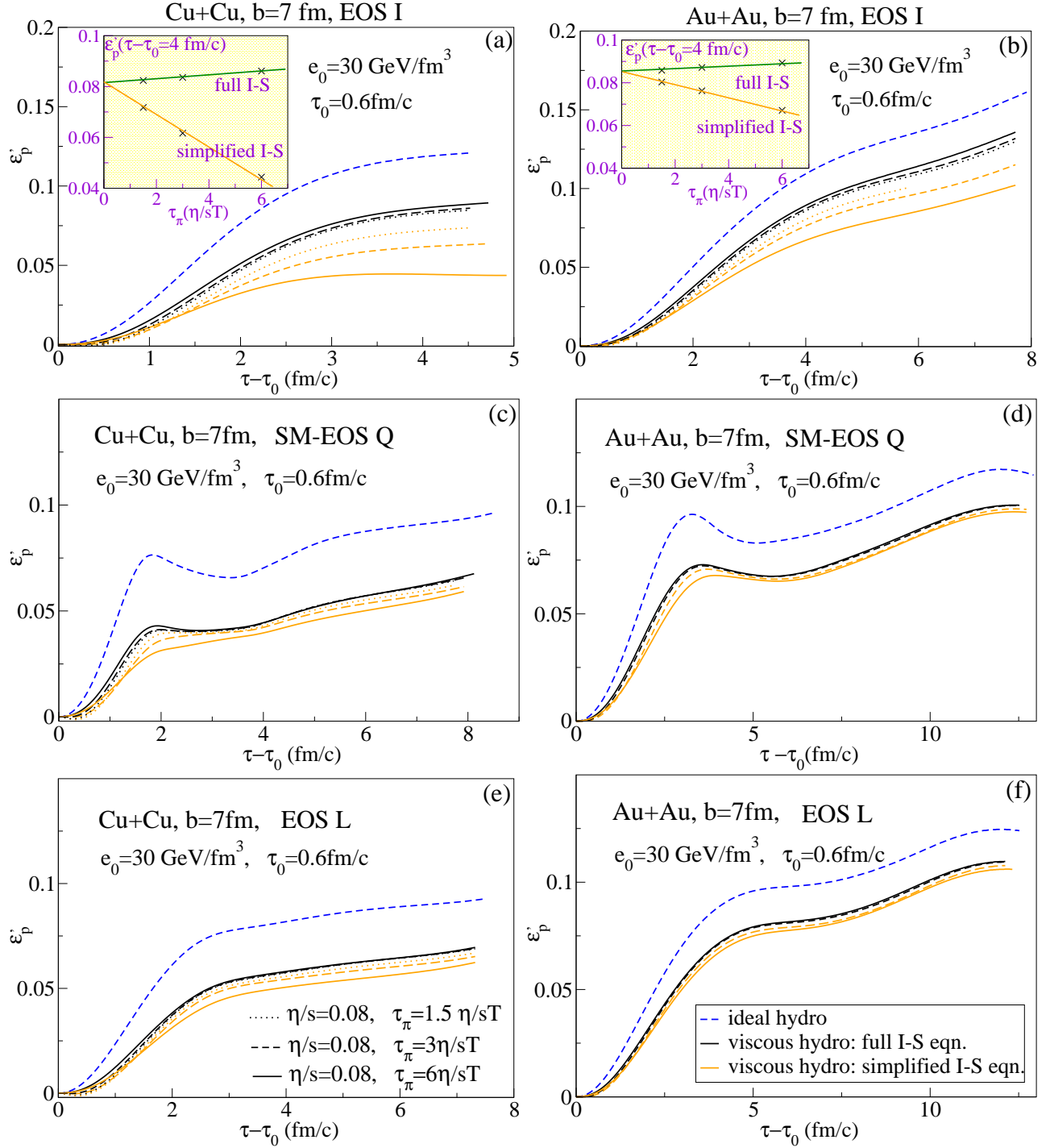

\includegraphics[width =0.47\linewidth,clip=]{Fig2a.eps}
\includegraphics[width =0.47\linewidth,clip=]{Fig2b.eps}\\
\includegraphics[width =0.47\linewidth,clip=]{Fig2c.eps}
\includegraphics[width =0.47\linewidth,clip=]{Fig2d.eps}\\
\includegraphics[width =0.47\linewidth,clip=]{Fig2e.eps}
\includegraphics[width =0.47\linewidth,clip=]{Fig2f.eps}
\caption{(Color online) Time evolution of the total momentum anisotropy 
$\varepsilon'_p$ for two collision systems (left: Cu+Cu; right: Au+Au), 
three equations of state (top: EOS~I; middle: SM-EOS~Q; bottom: EOS~L),
and three values of the kinetic relaxation time $\tau_\pi$ as indicated
(dotted, dashed and solid curves, respectively). The insets in the two
top panels show the $\tau_\pi$-dependence of the momentum anisotropy 
$\ecc_p$ at fixed time $\tau-\tau_0=4$\,fm/$c$. See text for discussion.
\label{F2}
}
\end{figure*}
%%%%%%%%%%%%%%%%%%%%%%%%%%%%%%%%%%%%%%%%%%%%%%%%%%%%%%%%%%%%%%%%%%%%%%%%%%%%%%
%

%
%%%%%%%%%%%%%%%%%%%%%%%%%%%%%% Fig. 3 v2(p_T) %%%%%%%%%%%%%%%%%%%%%%%%%%%%%%%%
\begin{figure*}[htbp]
\includegraphics[width=\linewidth,clip=]{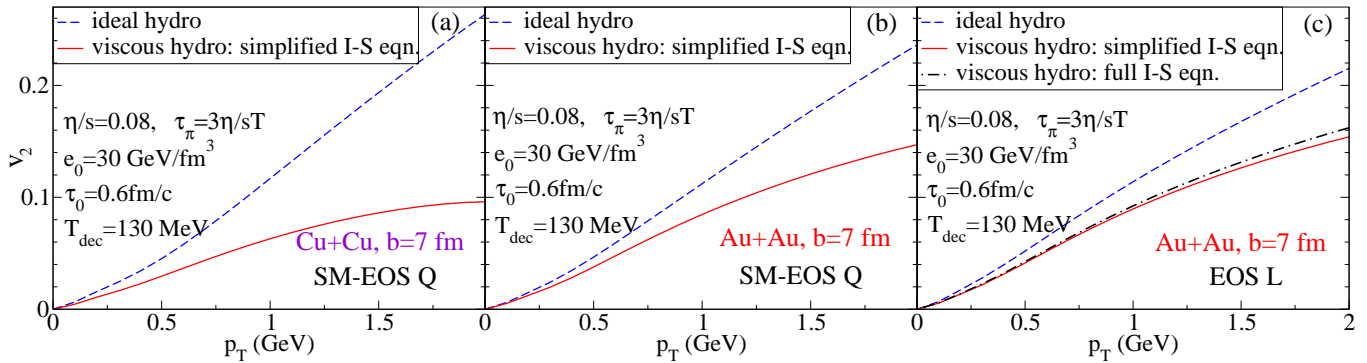}
\caption{(Color online) Differential elliptic flow $v_2(p_T)$ for
directly emitted pions (i.e. without resonance decay contributions),
comparing results for different collisions systems and equations
of state. (a) Cu+Cu at $b=7$\,fm with SM-EOS~Q. (b) Au+Au at 
$b=7$\,fm with SM-EOS~Q. (c) Au+Au at $b=7$\,fm with EOS~L.
Dashed (solid) lines correspond to ideal (viscous) fluid dynamics,
with parameters as indicated.
\label{F3}
}
\end{figure*}
%%%%%%%%%%%%%%%%%%%%%%%%%%%%%%%%%%%%%%%%%%%%%%%%%%%%%%%%%%%%%%%%%%%%%%%%%%%%%%
%

Figure~\ref{F2} shows the temporal evolution of the total momentum 
anisotropy $\ecc_p=\frac{\langle T^{xx}{-}T^{yy}\rangle}
{\langle T^{xx}{+}T^{yy}\rangle}$ averaged over the transverse plane%
\footnote{Note that $\ecc_p$ as defined here includes the effects from 
both flow velocity and shear pressure anisotropies \cite{Song:2007fn}.
In Ref.~\cite{Song:2007fn} we denoted it by $\ecc'_p$ in order to 
distinguish it from the flow-induced momentum anisotropy $\frac{
\langle T_0^{xx}{-}T_0^{yy}\rangle}{\langle T_0^{xx}{+}T_0^{yy}\rangle}$
which is based only on the ideal fluid part of the energy momentum tensor
and neglects anisotropies in the local fluid rest frame caused by the
shear pressure tensor $\pi^{mn}$. In the present work we drop the prime 
for convenience.}
for two collision systems (Cu+Cu at $b=7$\,fm on the left, Au+Au at 
$b=7$\,fm on the right) and three equations of state (EOS~I (top), 
SM-EOS~Q (middle), and EOS~L (bottom)). The blue dashed lines at the 
top indicate the result from ideal fluid dynamics, the black and orange 
lines below show viscous fluid dynamical results. The black lines show 
solutions of the full I-S equations, the orange ones for the simplified
I-S approach; in each case several values of the kinetic relaxation 
time $\tau_\pi$ are explored. Note that our full I-S equations 
(\ref{EQpimn}) do not use the identity (\ref{confid}) used in 
\cite{Romatschke:2007mq} which strictly holds only for conformal fluids 
(i.e. for the case of EOS~I in Fig.~\ref{F2}). We have, however, tested 
the two expressions on the left and right of Eq.~(\ref{confid}) against 
each other also for the other two equations of state (SM-EOS~Q and EOS~L) 
which are not conformally invariant, and found no discernible differences. 
Only for a very long relaxation time $\tau_\pi=12\eta/sT$ (not shown in 
Fig.~\ref{F2}) did we see for EOS~L a difference larger than the line 
width, with our result for $\ecc_p$ lying slightly above the one obtained 
with the conformal approximation (\ref{confid}). 

Comparison of the black and red lines in Fig.~\ref{F2} shows that the 
sensitivity of the momentum anisotropy $\ecc_p$ to the relaxation time 
$\tau_\pi$ is significantly larger for the simplified I-S equations (red)
than for the full I-S equations, and that the $\tau_\pi$-dependence of 
$\ecc_p$ even has the opposite sign for the two sets of equations.
With the full I-S equations, $\ecc_p$ moves slowly towards the ideal 
fluid limit as $\tau_\pi$ increases whereas with the simplified I-S 
equations $\ecc_p$ moves away from the ideal fluid limit, at a more 
rapid rate, resulting in a larger viscous suppression of the momentum
anisotropy. In the limit $\tau_\pi\to0$, both formulations approach the
same Navier-Stokes limit. The difference between full and simplified I-S 
equations is largest for EOS~I which is the stiffest of the three studied
equations of state, causing the most rapid expansion of the fireball. 
For this EOS, the simplified I-S equations allow for the largest 
excursions of $\pi^{mn}$ away from its Navier-Stokes limit, causing a
significant and strongly $\tau_\pi$-dependent increase of all viscous 
effects, including the suppression of the momentum anisotropy (Fig.~\ref{F2}) 
and elliptic flow (see Fig.~\ref{F4} below) and the amount of viscous 
entropy production (see Sec.~\ref{sec6}). 

For the other two equations of state, SM-EOS~Q and EOS~L, the difference 
between full and simplified I-S dynamics is much smaller, ranging from 
$\sim 5\%$ for Au+Au to $\sim 15\%$ for Cu+Cu for the largest $\tau_\pi$ 
value of $6\eta/sT$ studied here. Note that the viscous suppression of 
$\ecc_p$ is much stronger for the smaller Cu+Cu collision system than 
for Au+Au. For SM-EOS~Q and EOS~L (which yield rather similar results 
for $\ecc_p$, with differences not exceeding $\sim 10\%$), the results
from the full I-S equations (black lines) are almost completely independent
of $\tau_\pi$, even for the small Cu+Cu system. 

The insets in the two upper panels of Fig.~\ref{F2} illustrate the 
different $\tau_\pi$-dependences for $\ecc_p$ in the full and simplified
I-S formulations, by plotting the value of $\ecc_p$ for EOS~I at a fixed
time $\tau-\tau_0=4$\,dm/$c$ as a function of $\tau_\pi$. One sees that,
for the investigated range of relaxation times, the $\tau_\pi$-dependence 
is linear, but that the slope has different signs for the full and 
simplified I-S equations and is much smaller for the full I-S system. Even
though {\tt VISH2+1} cannot be run for much smaller $\tau_\pi$ values, due to
numerical instabilities that develop as the Navier-Stokes limit $\tau_\pi=0$
is approached, the lines corresponding to the full and the simplified I-S 
equations are seen to nicely extrapolate to the same Navier-Stokes point,
as they should. For SM-EOS~Q and EOS~L, the corresponding lines may no 
longer be linear, due to phase transition effects, but are still 
characterized by opposite slopes for the simplified and full I-S 
approaches, with almost vanishing slope in the full I-S case. This 
agrees with findings reported in \cite{Romatschke:2007mq,Luzum:2008cw}.
  
In Fig.~\ref{F3} the effects of changing the system size, EOS, and form of
I-S equations on the differential elliptic flow $v_2(p_T)$ for directly 
emitted pions is shown. The largest viscous suppression of elliptic flow
(by almost 70\% below the ideal fluid value at $p_T=2$\,GeV/$c$) is seen 
for the small Cu+Cu system, evolved with SM-EOS~Q and the simplified I-S 
equation. This is the result reported by us in \cite{Song:2007fn}. The 
middle panel of Fig.~\ref{F3} shows that this large $v_2$ suppression
is almost cut in half by going from Cu+Cu to Au+Au, the system studied
in \cite{Romatschke:2007mq}, even without modifying the EOS or the form 
of the I-S equations. Changing the EOS from SM-EOS~Q \cite{Song:2007fn} 
to EOS~L (which is close to the one used in \cite{Romatschke:2007mq}) 
reduces the viscous $v_2$ suppression by another quarter, from about 40\% 
to less than 30\% below the ideal fluid limit at $p_T=2$. Finally, 
replacing the simplified I-S equations used in \cite{Song:2007fn} by the 
full I-S equations employed by Romatschke \cite{Romatschke:2007mq} further 
reduces the suppression from about 28\% below the ideal fluid to 
$\sim 25\%$ at $p_T=2$\,GeV/$c$. This is consistent with the results
obtained \cite{Romatschke:2007mq}.

We conclude that the biggest contribution to the large difference between
the results reported in Refs. \cite{Song:2007fn} and \cite{Romatschke:2007mq}
arises from the different collisions systems studied, with much larger
viscous effects seen in the smaller Cu+Cu system than in Au+Au collisions.
The next most important sensitivity is to the EOS; for the most realistic 
EOS studied here, EOS~L, the differences between using the full or 
simplified I-S equations with $\tau_\pi=3\eta/sT$ are only about 10\% on 
a relative scale, or about 3\% on the absolute scale set by the elliptic 
flow from ideal fluid dynamics. For smaller $\tau_\pi$, this last 
difference would shrink even further.

The sensitivity to details of the EOS documented by the middle and right 
panels of Fig.~\ref{F3} gives an idea of how well one needs to know the
EOS if one wants to extract the specific shear viscosity $\eta/s$ from
experimental data using viscous hydrodynamics. One might argue that the 
difference between a first order phase transition implemented through 
SM-EOS~Q and a smooth crossover as in EOS~L should be sufficiently 
extreme to cover the maximal theoretical uncertainty. In this case,
Fig~\ref{F3} tells us that the maximal theoretical uncertainty on the
viscous suppression of $v_2$ (and therefore on $\eta/s$) should be about
25-30\%. This should be compared to the theoretical error introduced by our
present uncertainty of the initial spatial source eccentricity $\ecc$:
$\ecc$ differs by about 30\% between initializations based on the Glauber 
and color glass condensate (CGC) models \cite{Hirano:2005xf,Luzum:2008cw,%
Drescher:2006pi,Lappi:2006xc}, resulting in a $\sim30\%$ uncertainty of
the total magnitude of the elliptic flow in ideal fluid dynamics. We 
further caution that recent discussions about the value of the critical 
temperature for the quark-hadron transition \cite{Aoki:2006br,Cheng:2006qk} 
introduce an additional moment of uncertainty which is perhaps not covered 
by the range between SM-EOS~Q and EOS~L studied here. Therefore, while we 
agree with the authors of Ref.~\cite{Luzum:2008cw} that the uncertainty 
about the initial source eccentricity dominates over uncertainties related 
to different implementations of the I-S formalism, we think that the EOS 
should not be discounted prematurely as a possible source of significant 
additional theoretical uncertainty in the extraction of $\eta/s$.   

%
%%%%%%%%%%%%%%%%%%%%%%% Fig. 4 v2/ecc EOS I %%%%%%%%%%%%%%%%%%%%%%%%%%%%%%%%%
\begin{figure*}[thbp]
\includegraphics[width =\linewidth,clip=]{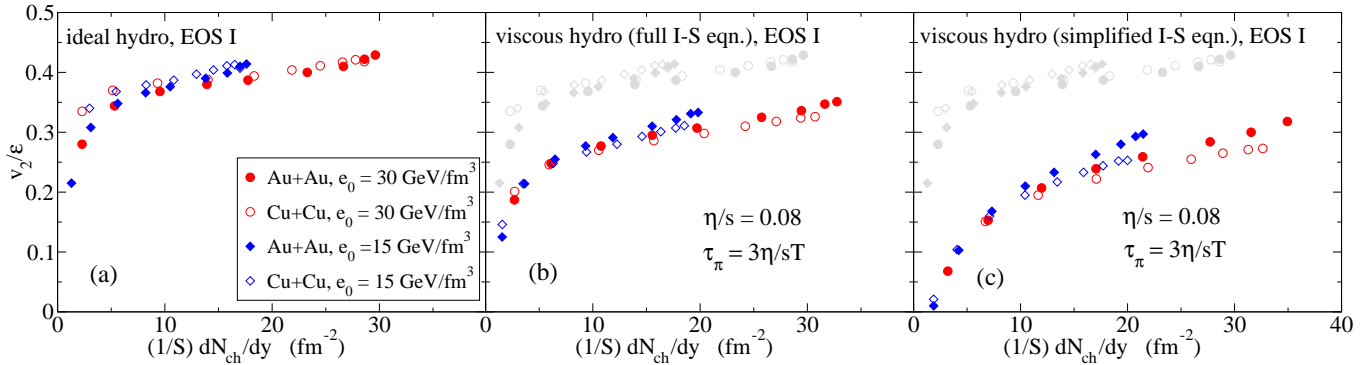}
\caption{(Color online) The eccentricity-scaled elliptic flow 
$v_2/\varepsilon$ as a function of charged multiplicity density,
$(1/S)\dNdy$, for a conformal fluid with EOS~I. Results for Cu+Cu and Au+Au
collisions with two different initial energy densities at a variety of
impact parameters, as indicated in the legend, are superimposed. 
Results from ideal fluid dynamics (a) are compared with
those from viscous hydrodynamics, using the full (b) and 
the simplified (c) Israel-Stewart equations, respectively.
In all cases approximate, but not perfect multiplicity scaling is
observed (see text for discussion). In panels (b) and (c),
the ideal fluid results from the left panel are reproduced as gray
symbols for comparison.
\label{F4}
}
\end{figure*}
%%%%%%%%%%%%%%%%%%%%%%%%%%%%%%%%%%%%%%%%%%%%%%%%%%%%%%%%%%%%%%%%%%%%%%%%%%%%%
%
 
%%%%%%%%%%%%%%%%%%%%%%%%%%%%%%%%%%%%%%%%%%%%%%%%%%%%%%%%%%%%%%%%%%%%%%%%%%%%%%
\section{Multiplicity scaling of $\bm{v_2/\ecc}$ in ideal and viscous 
hydrodynamics}
\label{sec5}
%%%%%%%%%%%%%%%%%%%%%%%%%%%%%%%%%%%%%%%%%%%%%%%%%%%%%%%%%%%%%%%%%%%%%%%%%%%%%%

In this section we explore the multiplicity scaling (as defined in the 
Introduction) of the eccentricity-scaled elliptic flow $v_2/\ecc$, comparing
ideal fluid dynamics with that of near-minimally viscous fluids with specific
shear viscosity $\frac{\eta}{s} ={\cal O}\left(\frac{1}{4\pi}\right)$. 

%%%%%%%%%%%%%%%%%%%%%%%%%%%%%%%%%%%%%%%%%%%%%%%%%%%%%%%%%%%%%%%%%%%%%%%%%%%%%%
\subsection{EOS~I: conformal fluids with $\bm{e=3p}$}
\label{sec5a}
%%%%%%%%%%%%%%%%%%%%%%%%%%%%%%%%%%%%%%%%%%%%%%%%%%%%%%%%%%%%%%%%%%%%%%%%%%%%%%

We begin with the simple case of a conformal fluid with the equation of 
state $e=3p$ (EOS~I), without phase transition. In this case the speed of 
sound is a constant, independent of temperature $T$, $c_s^2=\frac{1}{3}$. 
For the ideal fluid case, naive scaling arguments based on the scale 
invariance of the ideal fluid equations of motion would thus predict a 
constant $v_2/\ecc$, independent of multiplicity density $(1/S)\dNdy$. 
(The nuclear overlap area $S$ is computed as $S=\pi \sqrt{\La x^2\Ra 
\La y^2 \Ra}$ where $\La\dots\Ra$ denotes the energy density weighted 
average over the transverse plane.) The left panel in Fig.~\ref{F4} 
clearly contradicts this expectation. Freeze-out at $\Tdec=130$\,MeV 
cuts the hydrodynamic evolution of the momentum anisotropy $\ecc_p$
short before the elliptic flow has fully saturated. As the left panel
of the figure shows, this not only causes a strong suppression of $v_2/\ecc$
at low multiplicity densities, where the time between beginning of the 
hydrodynamic expansion and freeze-out becomes very short, but it also
breaks the multiplicity scaling at high multiplicity density, albeit more
weakly. At a fixed value of $(1/S)\dNdy$, one sees larger $v_2/\ecc$
for more central collisions initiated at lower collision energies
(corresponding to smaller $e_0$ parameters) than for more peripheral
collisions between the same nuclei at higher beam energies, and also 
for more central Cu+Cu collisions (with a rounder shape) than for more 
peripheral Au+Au collisions (with a more deformed initial shape). We 
find that the larger $v_2/\ecc$ values can be traced directly to 
somewhat longer lifetimes of the corresponding fireballs, i.e. to
the availability of more time to approach the saturation values of 
the momentum anisotropy and elliptic flow before reaching freeze-out. 
These freeze-out induced scaling violations in ideal fluid dynamics 
disappear at sufficiently high collision energies (i.e. large $e_0$) 
where the momentum anisotropy has time to fully saturate in {\em all} 
collision systems and at {\em all} impact parameters, before freezing 
out.  

The middle and right panels in Fig.~\ref{F4} show the ana\-lo\-gous results
for a minimally viscous fluid with $\frac{\eta}{s}=\frac{1}{4\pi}$ and
kinetic relaxation time $\tau_\pi=\frac{3\eta}{sT}$. Consistent with the 
discussion in the preceding section, the viscous suppression of the
elliptic flow is seen to be stronger if the simplified I-S equations
are used (right panel) than for the full I-S equation. (Although not 
shown, the curves in the right panel also show stronger sensitivity to 
the value of $\tau_\pi$ than those in the middle panel.) Along with the 
suppression of $v_2/\ecc$ by shear viscosity we see the appearance of 
scale-breaking effects that increase in proportion to the overall
suppression of elliptic flow: they are larger in the right than in the
middle panel. Shear viscosity breaks the multiplicity scaling of 
$v_2/\ecc$ because (as shown in the preceding section) viscous effects 
are larger in smaller collision fireballs. Consequently, if we compare 
different collision systems that produce the same charged multiplicity 
density $(1/S)\dNdy$, we find smaller $v_2/\ecc$ for Cu+Cu than for 
Au+Au collisions, and for peripheral Au+Au collisions at higher collision 
energy than for more central Au+Au collisions at lower collision energy. 

Viscous effects also generate entropy, i.e. they increase the final 
charged multiplicity $\dNdy$. Comparing in the middle and right panels of 
Fig.~\ref{F4} the gray (shaded) symbols from ideal fluid dynamics with 
the colored (solid) symbols for viscous hydrodynamics, points corresponding 
to the same collision system and impact parameter are seen to be shifted 
to the right. This enhances the scaling violations:
for a given collision system, impact parameter and collision energy, 
viscosity decreases the eccentricity scaled elliptic flow $v_2/\ecc$,
pushing the corresponding point downward in the diagram, and
simultaneously increases the entropy, pushing the corresponding point
horizontally to the right. The combination of these two effects separates
the curves for different collision systems and energies farther in
viscous hydrodynamics than in ideal fluid dynamics.  

%
%%%%%%%%%%%%%%%%%%% Fig. 5 v2/ecc EOSQ and EOS L %%%%%%%%%%%%%%%%%%%%%%%%%%%%
\begin{figure*}[thbp]
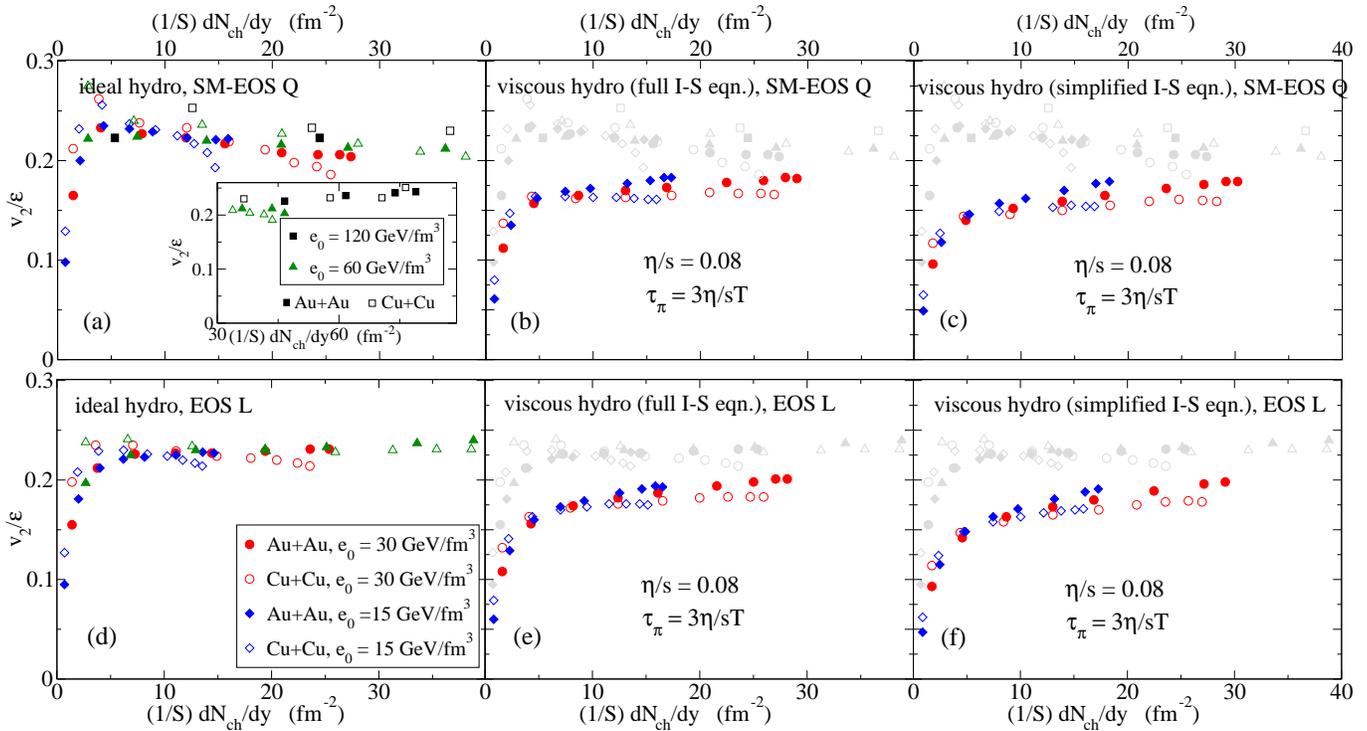

\includegraphics[width =\linewidth,clip=]{Fig5a.eps}\\
\includegraphics[width =\linewidth,clip=]{Fig5b.eps}
\caption{(Color online) Same as Fig.~\ref{F4}, but for SM-EOS~Q (top
row) and EOS~L (bottom row). For the ideal fluid case (a,d)
an extended range of $e_0$ values up to $e_0=120$\,GeV/fm$^3$ was
studied, in order to show that $v_2/\ecc$ eventually increases again
at higher collision energies \cite{Kolb:1999it}.
\label{F5}
}
\end{figure*}
%%%%%%%%%%%%%%%%%%%%%%%%%%%%%%%%%%%%%%%%%%%%%%%%%%%%%%%%%%^%%%%%%%%%%%%%%%%%%
%

%%%%%%%%%%%%%%%%%%%%%%%%%%%%%%%%%%%%%%%%%%%%%%%%%%%%%%%%%%%%%%%%%%%%%%%%%%%%%%
\subsection{Phase transition effects: EOS~Q and EOS~L}
\label{sec5b}
%%%%%%%%%%%%%%%%%%%%%%%%%%%%%%%%%%%%%%%%%%%%%%%%%%%%%%%%%%%%%%%%%%%%%%%%%%%%%%

Figure~\ref{F5} shows the analogous results if the fluid evolves under the
influence of an equation of state with a quark-hadron phase transition, 
EOS~Q (top row) or EOS~L (bottom row). Again approximate multiplicity
scaling of $v_2/\ecc$ is observed, but small scale-breaking effects are 
visible in both ideal and viscous hydrodynamics. For the equations of 
state with a phase transition, the scale-breaking effects are actually 
larger in the ideal than in the viscous case, i.e. {\em in
viscous hydrodynamics $v_2/\ecc$ shows better multiplicity scaling
than in ideal fluid dynamics!} We interpret the large scale-breaking 
effects in the ideal fluid case as a complication arising from 
interference between the freeze-out process and the weak acceleration of 
matter in the phase transition region. This interpretation is supported
by a comparison between SM-EOS~Q with its first-order phase transition
(upper left panel in Fig.~\ref{F5}) and the smooth crossover transition 
in EOS~L (lower left panel): for ideal fluids, the scale-breaking effects 
are obviously larger for SM-EOS~Q than for EOS~L. As already observed in 
\cite{Song:2007fn}, shear viscosity effectively smears out the phase 
transition and reduces its effect on the dynamics. In Fig.~\ref{F5} this 
is clearly seen on the left side of each panel (i.e. at small values of 
$\frac{1}{S}\frac{dN_\mathrm{ch}}{dy}$) where for the ideal fluid 
$v_2/\ecc$ shows a non-monotonic peak structure \cite{Kolb:1999it} that 
is completely gone in the viscous case.

Comparing Figs.~\ref{F4} and \ref{F5}, we see much smaller differences
between the full (middle panels) and simplified I-S equations (right
panels) for SM-EOS~Q and EOS~L than for EOS~I. This is consistent
with our observations in Sec.~\ref{sec4} where the largest differences
between full and simplified I-S equations was also seen for the rapidly
evolving fireballs whose expansion is pushed by the very stiff EOS~I. 

It is interesting to observe that, for ideal fluids, EOS~L leads to about
10\% more elliptic flow under RHIC conditions than SM-EOS~Q. The reason
is that in the phase transition region EOS~L is stiffer than SM-EOS~Q. 
This plays an important role at RHIC because the softness of the EOS near
$T_c$ inhibits the buildup of elliptic flow exactly under RHIC conditions
\cite{Kolb:1999it}. As a corollary we note that, if RHIC elliptic flow 
data exhaust ideal fluid predictions made with SM-EOS~Q \cite{reviews}, 
they will not exhaust ideal fluid predictions based on EOS~L, thus 
leaving some room for shear viscous effects.

%%%%%%%%%%%%%%%%%%%%%%%%%%%%%%%%%%%%%%%%%%%%%%%%%%%%%%%%%%%%%%%%%%%%%%%%%%%%%%
\subsection{Viscous suppression of $\bm{v_2}$: systematics}
\label{sec5c}
%%%%%%%%%%%%%%%%%%%%%%%%%%%%%%%%%%%%%%%%%%%%%%%%%%%%%%%%%%%%%%%%%%%%%%%%%%%%%%

Even at the highest collision energies (or $e_0$ values) studied 
in Figs.~\ref{F4} and \ref{F5}, the slope of $v_2/\ecc$ as a function of 
$\frac{1}{S}\frac{dN_\mathrm{ch}}{dy}$ remains positive, i.e. $v_2/\ecc$ 
continues to increase and evolve in direction of the asymptotic ideal 
fluid limit. This implies that at higher collision energies the 
importance of viscous effects decreases. This observations parallels the 
one made in \cite{Song:2007fn}, namely that with increasing collision 
energy the $p_T$ range increases over which viscous hydrodynamic predictions
for the single-particle momentum spectra can be trusted. The reason is
in both cases that with increasing collision energy the time until
freeze-out grows, and that (at least for constant $\eta/s$ as assumed
here and in \cite{Song:2007fn}) during the later stages of the expansion 
shear viscous effects are small.

%
%%%%%%%%%%%%%%%%%%%%%%%%%%%%%% Fig. 6 v2 supp %%%%%%%%%%%%%%%%%%%%%%%%%%%%%%%
\begin{figure}[ht]
\includegraphics[width=\linewidth,clip=]{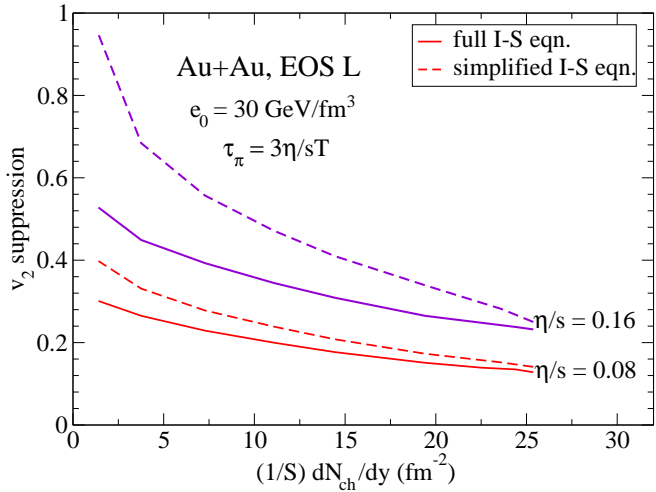}
\caption{(Color online) Viscous suppression of elliptic flow,
$(v_2^\mathrm{ideal}-v_2^\mathrm{viscous})/v_2^\mathrm{ideal}$,
as a function of $(1/S)\dNdy$ for Au+Au with EOS~L, $\tau_\pi=3\eta/sT$ 
and two values of $\eta/s$ as indicated. Solid (dashed) lines correspond 
to using the full (simplified) I-S equations, respectively. 
\label{F6}
}
\end{figure}
%%%%%%%%%%%%%%%%%%%%%%%%%%%%%%%%%%%%%%%%%%%%%%%%%%%%%%%%%%%%%%%%%%%%%%%%%%
%

Figure~\ref{F6} shows this more quantitatively. We plot the fractional
decrease of the elliptic flow relative to its ideal fluid dynamical value,
$(v_2^\mathrm{ideal}-v_2^\mathrm{viscous})/v_2^\mathrm{ideal}$,
as a function of multiplicity density. Larger multiplicity densities
lead to smaller viscous suppression effects. Larger viscosity results
in stronger suppression of the elliptic flow. The suppression effects
are weaker if the full I-S equations are used than in the simplified
approach of Ref.~\cite{Song:2007fn} (which, as discussed in Sec.~\ref{sec4},
also suffers from strong sensitivity to $\tau_\pi$). For minimal
viscosity, $\eta/s=1/4\pi$, the $p_T$-integrated elliptic flow $v_2$
in Au+Au collisions at RHIC is suppressed by about 20\%. The suppression 
is larger at lower energies but will be less at the LHC. 

%%%%%%%%%%%%%%%%%%%%%%%%%%%%%%%%%%%%%%%%%%%%%%%%%%%%%%%%%%%%%%%%%%%%%%%%
\subsection {A look at experimental data}
\label{sec5d}
%%%%%%%%%%%%%%%%%%%%%%%%%%%%%%%%%%%%%%%%%%%%%%%%%%%%%%%%%%%%%%%%%%%%%%%%

%
%%%%%%%%%%%%%%%%%%%%%%%%%%%%%% Fig. 7 %%%%%%%%%%%%%%%%%%%%%%%%%%%%%%%%%%
\begin{figure*}[ht]
\includegraphics[bb=17 20 510 440,width=0.46\linewidth,clip=]%
                {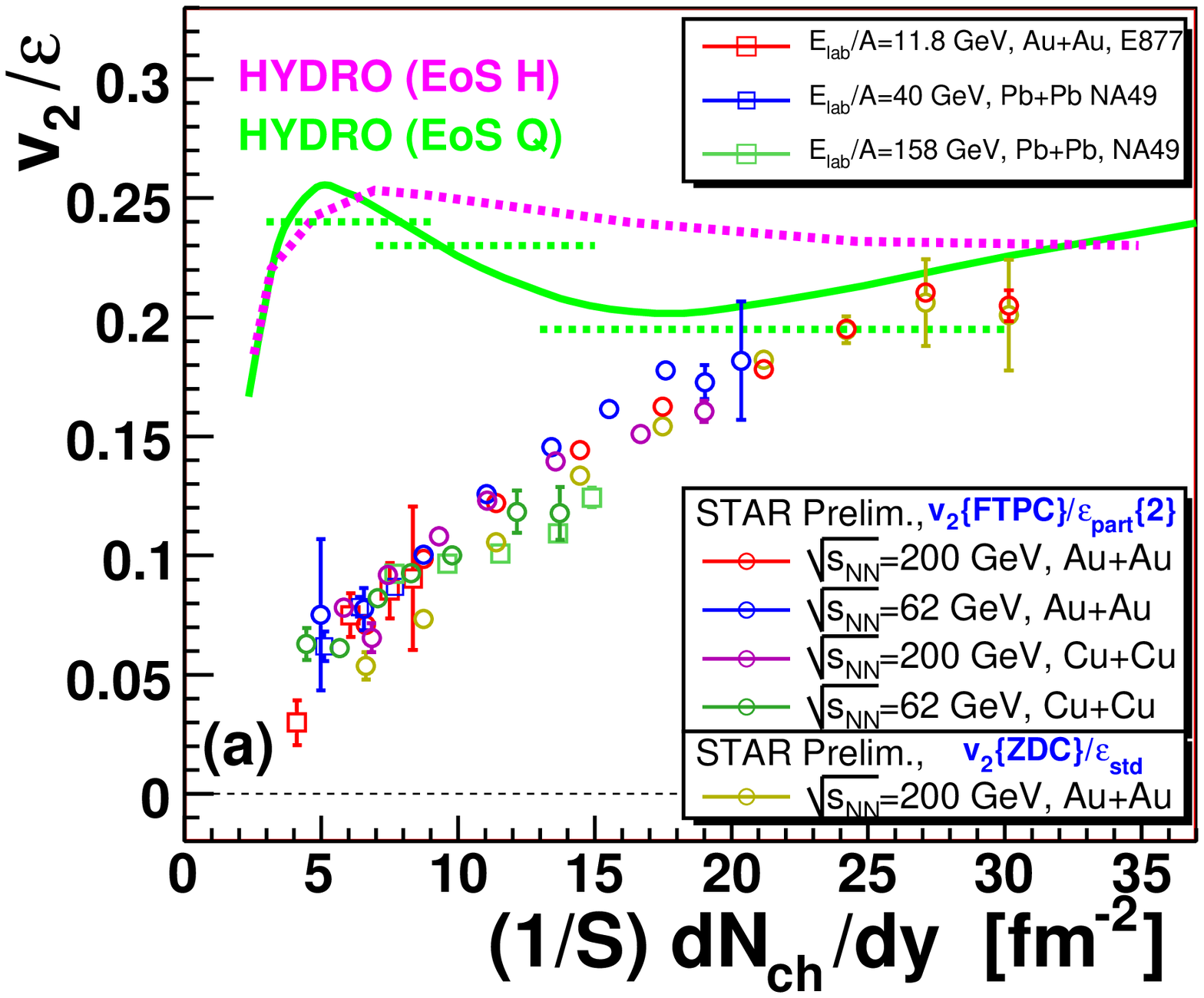}
\includegraphics[width=0.49\linewidth,height=7.05cm,clip=]%
                {Fig7b.eps}
\caption{(Color online) (a) The experimental observation
of multiplicity scaling for $v_2/\ecc$, with data from Au+Au, Pb+Pb,
and Cu+Cu collisions at RHIC, SPS and AGS \cite{VoloshinQM06}.
(b) Theoretical prediction of approximate multiplicity scaling 
from viscous hydrodynamics using the full I-S equations, for three 
different (constant) specific entropy values $\eta/s=0.08,\,0.16,\,0.24$. 
\label{F7}
}
\end{figure*}
%%%%%%%%%%%%%%%%%%%%%%%%%%%%%%%%%%%%%%%%%%%%%%%%%%%%%%%%%%%%%%%%%%%%%%%%%%
%

Figure~\ref{F7}(a) shows the famous experimental plot by
Voloshin \cite{VoloshinQM06} with provides empirical evidence for
multiplicity scaling of $v_2/\ecc$. The lines labelled ``HYDRO'' are
sketches for expectations from ideal fluid dynamics, based on the
calculations presented in \cite{Kolb:1999it} for $v_2$ in Au+Au collisions 
at fixed impact parameter $b=7$\,fm as a function of multiplicity 
(parametrized by $e_0$). They should be replaced by the curves shown
in the left panels of Fig.~\ref{F5}. 

In Fig.~\ref{F7}(b) we present multiplicity scaling curves for
$v_2/\ecc$ obtained from viscous hydrodynamics with the full I-S equations.
On a superficial level, the theoretical curves show qualitative similarity 
with the experimental data, giving correct ball-park numbers if one 
assumes $\eta/s \sim 0.24 \sim 3/4\pi$. Interestingly, ignoring 
experimental error bars, one can see evidence for small scaling 
violations in the experimental data whose pattern agrees with the 
theoretical predictions from viscous hydrodynamics (see discussion at 
the end of Sec.~\ref{sec5a}): the 62.5~$A$\,GeV Au+Au data lie slightly 
above the $200\,A$\,GeV Au+Au points, and the $200\,A$\,GeV Cu+Cu points
fall slightly below the $62.5\,A$\,GeV Au+Au data. Of course, these fine 
features of the experimental data are presently not statistically 
significant; much more precise data are needed to confirm or disprove 
the theoretical predictions, but upcoming high-statistic runs at RHIC 
should be able to deliver them.

Closer inspection of the two panels in Fig.~\ref{F7} shows, however, 
that the theoretical scaling curves have the wrong slope: on the left side
of the plot, i.e. for small multiplicity densities, the data seem to 
point towards larger specific shear viscosities $\frac{\eta}{s}>3\times
\frac{1}{4\pi}$ whereas on the right side of the plot, for 
$\frac{1}{S}\frac{dN_\mathrm{ch}}{dy} > 20$\,fm$^{-2}$, the experimental 
data require smaller shear viscosities, 
$\frac{\eta}{s}\lesssim(1{-}2)\times\frac{1}{4\pi}$. But this is not at
all unexpected: Collisions represented by points in the right half of the 
plot correspond to high collision energies and large initial energy
densities whose expanding fireballs spend the largest fraction of their 
life in the QGP phase. Fireballs created in collisions represented by 
points in the left part of the diagram have smaller initial energy densities
and thus spend most of their time in the much more viscous hadronic phase
\cite{Hirano:2005xf}. A meaningful comparison between theory and experiment
thus must necessarily account for the temperature dependence of $\eta/s$
and its dramatic increase during the quark-hadron phase transition
\cite{Hirano:2005wx}. This would lead to scaling curves in Fig.~\ref{F7}(b) 
with a larger slope that can better reproduce the data. What one can say 
already now is that the high-energy end of Fig.~\ref{F7} requires very 
small specific shear viscosity $\eta/s$ for the QGP, of the same order as 
the minimal value postulated in \cite{Ads-CFT,son-PRL05} (unless the 
initial source eccentricity $\ecc$ was strongly underestimated in the 
experimental data).   

%%%%%%%%%%%%%%%%%%%%%%%%%%%%%%%%%%%%%%%%%%%%%%%%%%%%%%%%%%%%%%%%%%%%%%%%%%%%
\section{Multiplicity scaling of entropy production in viscous hydrodynamics}
\label{sec6}
%%%%%%%%%%%%%%%%%%%%%%%%%%%%%%%%%%%%%%%%%%%%%%%%%%%%%%%%%%%%%%%%%%%%%%%%%%%%

In the absence of shock waves, ideal fluid dynamics conserves entropy.
Correspondingly, the final multiplicity per unit rapidity is directly 
determined by the total initial entropy per unit rapidity: $\frac{dN}{dy}
=\mathrm{const.}\times \tau_0 \int dx\, dy\, s(x,y;\tau_0)$. Numerical
discretization of the hydrodynamic evolution equations introduces a
small amount of ``numerical viscosity'', however, which can not be 
fully avoided. To minimize numerical viscosity effects, the flux-corrected 
transport algorithm SHASTA used in the numerical solution of both the 
ideal and viscous fluid equations \cite{AZHYDRO,Song:2007fn} employs 
an ``antidiffusion step'' involving a parameter called ``antidiffusion
constant'' \cite{Von-Gersdorff:1986yf}. Numerical viscosity effects 
are maximal if this parameter is set to zero. In all our simulations 
we used 0.125 for the antidiffusion constant \cite{Von-Gersdorff:1986yf}, 
resulting in about 0.3\% entropy production by numerical viscosity in 
the ideal fluid case. In comparison with the ${\cal O}(10{-}15\%)$ 
entropy production in a viscous fluid with minimal shear viscosity 
(see below) this can be neglected\footnote{%
Effects from numerical viscosity depend on the spacing
of the space-time grid used in the simulation. For the results presented
in this paper we used $\Delta x= \Delta y =0.1$\,fm and $\Delta \tau = 
0.04$\,fm/$c$. To check the effects of numerical viscosity in the ideal
fluid case we also performed simulations with AZHYDRO where 
we either set the antidiffusion constant to zero or increased and 
decreased the grid spacings $(\Delta x, \Delta y, \Delta\tau)$ by a 
factor 2 to 4. To maximize numerical viscosity effects, we used EOS~Q 
with a strong first order phase transition which generates shocks and 
associated large velocity gradients. We found that {\em decreasing} 
the grid spacing by a factor 2 has no visible effect on the 
average radial flow of the fluid in central Au+Au collisions but 
increases the momentum anisotropy of the ideal fluid in peripheral 
Au+Au collisions at $b=7$\,fm by $2{-}3\%$. A further reduction by 
another factor 2 doesn't even affect the momentum anisotropy any 
more, indicating that numerical viscosity effects have been basically
reduced to zero. If we {\em increase} the grid spacing by a factor two, 
the effects are a bit larger: the total entropy production by numerical 
viscosity increases by about 1\%, the average radial flow at $b=0$ 
changes by 0.5\%, and the momentum anisotropy experiences a relative 
suppression of about 5\%. The largest effects are seen when maximizing 
the numerical viscosity at fixed grid spacing by setting the 
antidiffusion constant to zero. In this case the average radial flow 
in central Au+Au collisions again changes only by 0.5\% (which is 
negligible compared to the strong increase in transverse acceleration 
that we see in the viscous fluid), but the momentum anisotropy is 
reduced by about 10\%. In the final pion $v_2$ this is reflected by 
a $p_T$-dependent reduction that increases with $p_T$, just like effects
from real shear viscosity \cite{Romatschke:2007mq,Chaudhuri:2007zm,%
Song:2007fn,Dusling:2007gi,Teaney:2003kp}; the $p_T$-integrated pion 
elliptic flow is reduced by less than 4\% even in this extreme case.
We conclude that numerical viscosity does not increase transverse 
acceleration but suppresses momentum anisotropy similar to real shear
viscosity. For the parameters used in this paper, and for the equations 
of state studied here which do not have a sharp phase transition, 
numerical viscosity effects on the elliptic flow do not exceed 
$1{-}2\%$ and can thus be neglected relative to effects from real 
shear viscosity. 
}.

Similar to the second paper of Ref.~\cite{Baier:2006gy}, we compute 
entropy production by exploiting the proportionality of final entropy
to final charged multiplicity. We compute the final multiplicity 
$\dNdy$ for both ideal and viscous hydrodynamics and then equate 
the fractional increase in $\dNdy$ with the fractional increase
in $d{\cal S}/dy$. This ignores a small negative correction due to
the viscous deviation of the distribution function on the freeze-out 
hypersurface from local equilibrium \cite{Israel:1976tn,Muronga:2006zw}
which slightly reduces the entropy per finally observed particle in the
viscous case. The real entropy production is thus slightly smaller
than calculated with our prescription. However, since on the freeze-out 
surface the viscous pressure components are small \cite{Song:2007fn}, 
this correction should be negligible. 

%
%%%%%%%%%%%%%%%%%%%%%%%%%%%%%% Fig. 8 %%%%%%%%%%%%%%%%%%%%%%%%%%%%%%%%%%
\begin{figure*}[htbp]
\includegraphics[bb=20 20 600 740,height =0.49\linewidth,angle=270,clip=]%
                {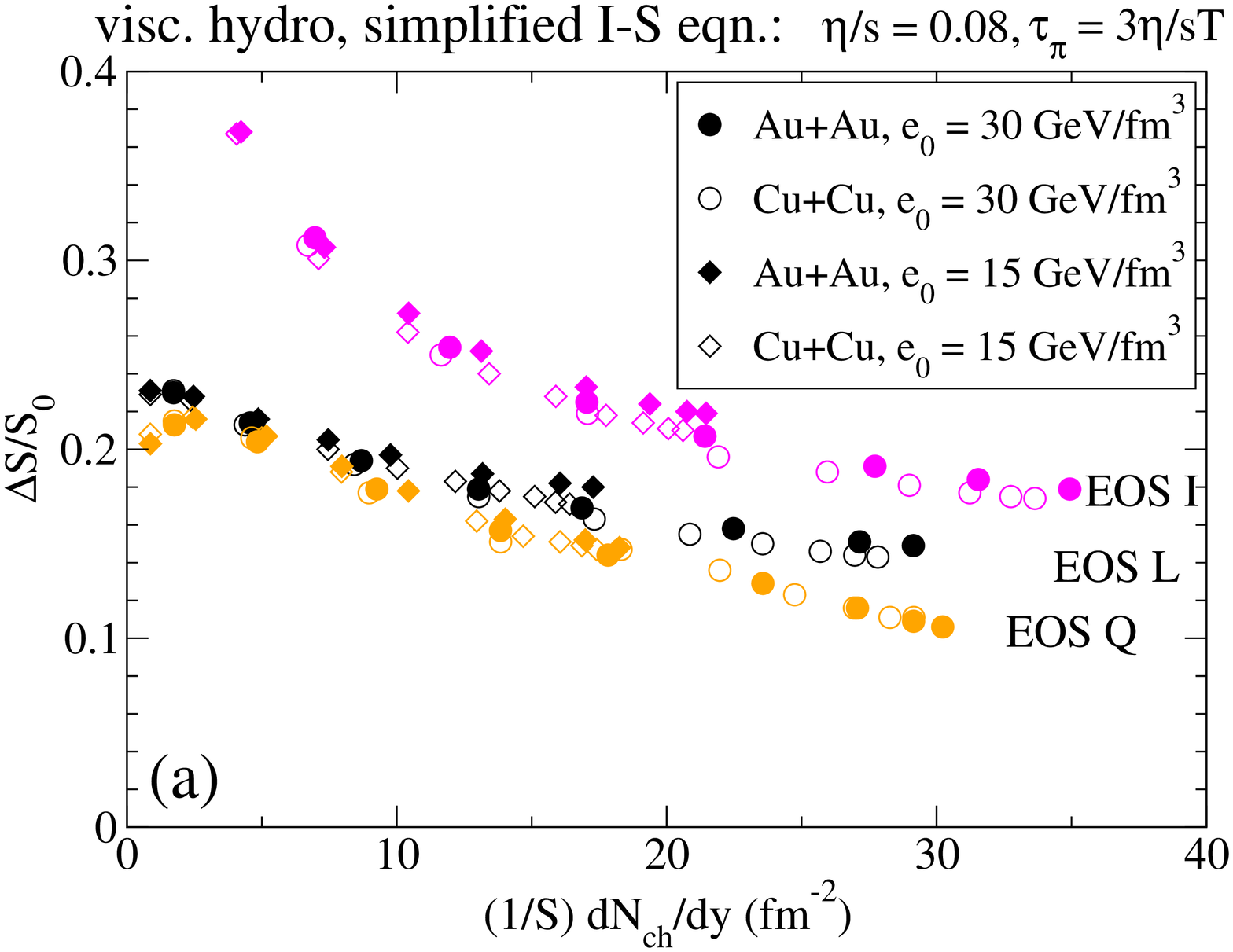}
\includegraphics[bb=20 20 600 740,height =0.49\linewidth,angle=270,clip=]%
                {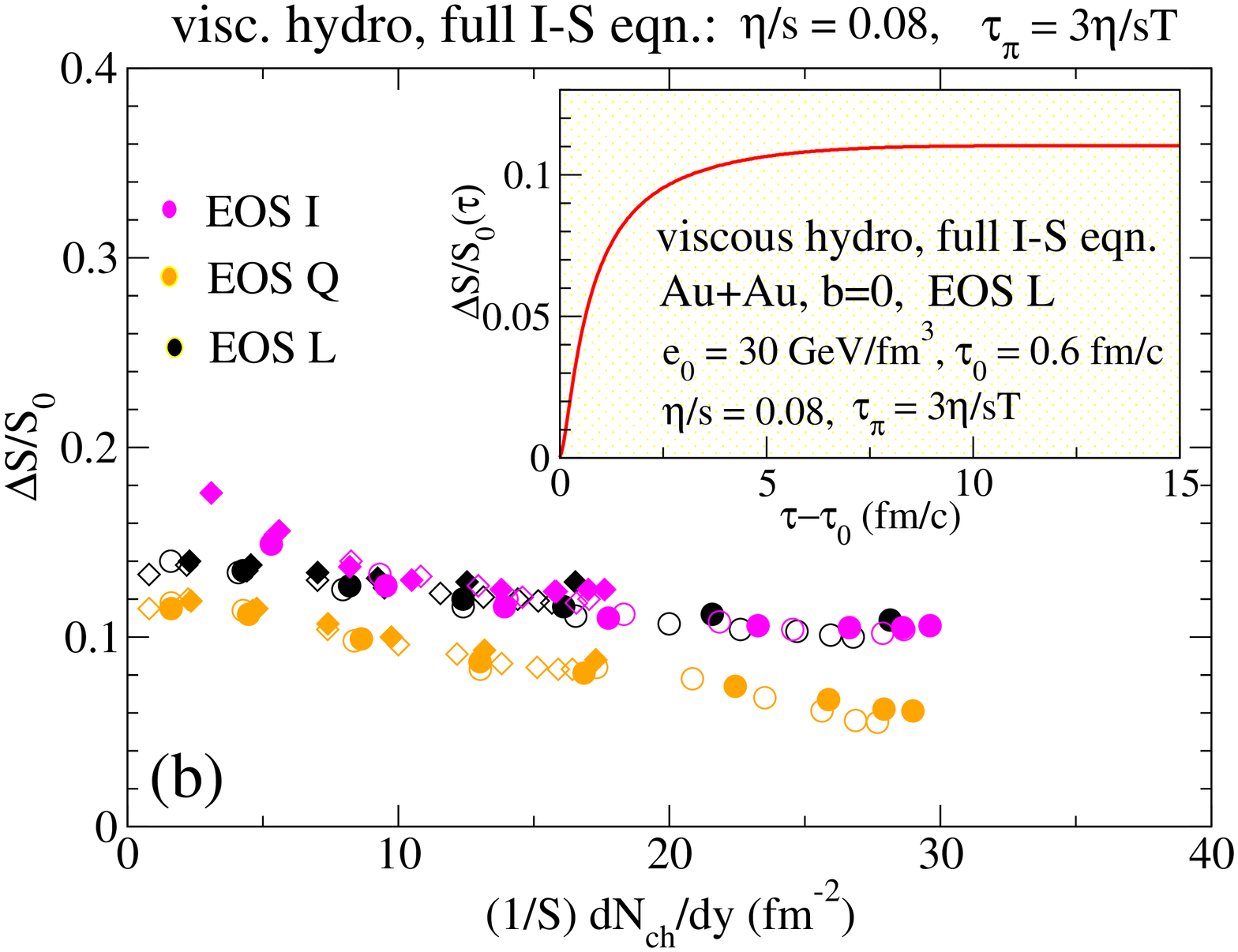}
\caption{(Color online) Entropy production $\Delta{\cal S}$, normalized
by the initial entropy ${\cal S}_0$, as a function of charged
multiplicity density $\frac{1}{S}\frac{dN_\mathrm{ch}}{dy}$. Calculations 
with {\tt VISH2+1} were performed for Au+Au and Cu+Cu collisions at various 
impact parameters and collision energies, using $\eta/s=0.08$, 
$\tau_\pi=3\eta/sT$, and three different equations of state (EOS~I, 
SM-EOS~Q, and EOS~L). (a) Simplified I-S equations. (b) Full I-S equations. 
The inset in panel (b) shows the entropy production as a function of time, 
for central Au+Au collisions with parameters as indicated in the legend.
\label{F8}
}
\end{figure*}
%%%%%%%%%%%%%%%%%%%%%%%%%%%%%%%%%%%%%%%%%%%%%%%%%%%%%%%%%%%%%%%%%%%%%%%%%%
%

We checked the above procedure by also directly integrating the viscous
entropy production rate $\partial\cdot s = \pi^{\mu\nu} \pi_{\mu\nu}/2\eta$
over the space-time volume enclosed between the initial condition Cauchy 
surface and the final freeze-out surface. This method results in slightly 
larger entropy production, the relative difference amounting to about 0.7\%
(or about 0.07\% in the absolute value of $\Delta{\cal S}/{\cal S}_0$) for 
central Au+Au collisions. Since the estimate from the final multiplicity 
gives a lower entropy production value even without accounting for the 
somewhat smaller entropy per particle in the viscous case, we conclude 
that entropy production due to {\em numerical} viscosity must be a bit 
smaller in the viscous fluid than in the ideal one. This is not 
unreasonable, given the observation in \cite{Song:2007fn} that, compared 
to the ideal fluid case, the physical viscosity smoothens the strong 
velocity gradients near the quark-hadron phase transition, thereby 
presumably also reducing the effects of numerical viscosity.

We note that our viscous evolution starts earlier (at $\tau_0=0.6$\,fm/$c$) 
than that of Ref.~\cite{Baier:2006gy} (who use $\tau_0=1$\,fm/$c$). This
earlier start results in larger entropy production fractions. As the inset 
in Fig.~\ref{F8}(b) shows, most of the entropy is produced during the
early stage of the expansion. We have confirmed that the difference between
Ref.~\cite{Baier:2006gy} and the work here is quantitatively reproduced 
by the entropy generated during the time interval from 0.6 to 1.0\,fm/$c$, 
which can be calculated to excellent approximation analytically 
\cite{Gyulassy85} (using Eq. (D3) in Ref.~\cite{Song:2007fn}) by assuming 
boost-invariant longitudinal expansion without transverse flow during this 
period.

Figure~\ref{F8} shows the viscous entropy production $\Delta{\cal S}$, as 
a fraction of the initial entropy ${\cal S}_0$, for Cu+Cu and Au+Au
collisions at various impact parameters and collision energies, as a
function of multiplicity density. One observes approximate multiplicity
scaling of the fractional entropy production, with scaling functions
that depend on the equation of state and, for non-zero kinetic relaxation
time, on the form of the Israel-Stewart equations used in the simulation.
As with $v_2/\ecc$ we see small scale-breaking effects, but generally the
produced entropy fraction shows better multiplicity scaling than elliptic 
flow. The scale breaking effects for the viscous entropy production rate 
go in the same direction as with elliptic flow insofar as, at the same 
value of $\frac{1}{S}\frac{dN_\mathrm{ch}}{dy}$, larger collision systems 
and more central collisions produce fractionally more entropy than smaller
or more peripheral collisions, due to their longer lifetimes before 
freeze-out.

%
%%%%%%%%%%%%%%%%%%%%%%%%%%%%%% Fig. 9 %%%%%%%%%%%%%%%%%%%%%%%%%%%%%%%%%%
\begin{figure}[hbt]
\includegraphics[bb=20 20 600 740,height =\linewidth,angle=270,clip=]{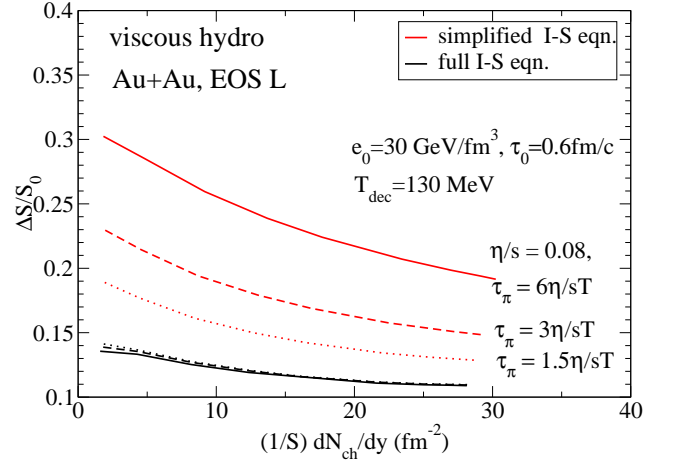}
\caption{(Color online) Sensitivity of the entropy production ratio 
$\Delta{\cal S}/{\cal S}_0$ shown in Fig.~\ref{F8} to the kinetic
relaxation time $\tau_\pi$, for the Au+Au collision system with 
$e_0=30$\,GeV/fm$^3$ (corresponding to a collision energy of 
$\sqrt{s}\approx200\,A$\,GeV). The three red curves (upper set)
are for the simplified Israel-Stewart equations, the three black 
curves (lower set) for the full I-S equations. Solutions with the
full I-S equations produce less entropy and show very little 
sensitivity to $\tau_\pi$.
\label{F9}
}
\end{figure}
%%%%%%%%%%%%%%%%%%%%%%%%%%%%%%%%%%%%%%%%%%%%%%%%%%%%%%%%%%%%%%%%%%%%%%%%%%
%

For $\tau_\pi=3\eta/sT$, Figure~\ref{F8} shows that the simplified I-S
equations (left panel) cause almost twice as much entropy production
as the full I-S system! Figure~\ref{F9} clarifies that, when the simplified
I-S equations are used, entropy production depends very sensitively on
the kinetic relaxation time $\tau_\pi$, approaching the much smaller
and almost completely $\tau_\pi$-independent entropy production rates
of the full I-S framework in the limit $\tau_\pi\to0$. The large amount
of extra entropy production for non-zero $\tau_\pi$ in the simplified
I-S approach must thus be considered as unphysical. This is important because
this artificial extra entropy production (caused by unphysically large
excursions of the viscous shear pressure tensor $\pi^{mn}$ away from
its Navier-Stokes value $\pi^{mn}=2\eta\sigma^{mn}$) manifests itself
as additional charged hadron multiplicity in the observed final state
(seen as a shift of all points in the left panel of Fig.~\ref{F8} 
towards larger values of $\frac{1}{S}\frac{dN_\mathrm{ch}}{dy}$). Since
the final multiplicity is used to normalize the initial energy density 
$e_0$, this causes a significant distortion of the initial conditions
corresponding to a given set of experimental data, affecting their
physical interpretation.

We conclude that using the full I-S equations is mandatory if one wants 
to minimize artificial effects of shear viscosity on entropy production
and elliptic flow in the realistic situation of non-zero kinetic relaxation
times. (We note that, while the value of $\tau_\pi$ for the QGP created at 
RHIC is presently unknown, it can obviously not be zero.) From the right 
panel in Fig.~\ref{F8} we see that this removes most of the large 
difference in entropy production between the rapidly exploding EOS~I
fireballs and their more leisurely expanding cousins that evolve under
the influence of EOS~Q or EOS~L. Still, even for the full I-S equations 
we see $15-25\%$ differences between the entropy production rates for 
EOS~Q (first order phase transition) and EOS~L (rapid crossover 
transition). The differences are largest for the most central Au+Au and 
Cu+Cu collisions at top RHIC energies. The somewhat stiffer nature of 
EOS~L near $T_c$ causes the fireball to expand faster and with higher 
acceleration, leading to larger viscous effects than for EOS~Q. The 
differences in entropy production caused by this variation of the EOS is 
of similar magnitude as its effect on the viscous suppression of 
$v_2/\ecc$ discussed at the end of Sec.~\ref{sec4}.

An important comment relates to the negative overall slope of the scaling
curves for entropy production shown in Figs.~\ref{F8} and \ref{F9}: Since
peripheral collisions produce relatively more entropy than central 
collisions, and the produced entropy is reflected in the final charged 
hadron multiplicity, the collision centrality dependence of hadron
multiplicities is altered by viscous effects. When viscous effects are
accounted for, the charged multiplicity $\dNdy$ will rise more slowly
as a function of the number of participant nucleons $N_\mathrm{part}$
than for an ideal fluid with the same set of initial conditions. In a
Glauber model parametrization of the initial conditions \cite{reviews} 
this tempering effect will have to be compensated for by increasing the 
``hard'' component in the initial entropy production. i.e. the component
that scales with the density of binary collisions and is thus responsible 
for the non-linear increase of $\dNdy$ with $N_\mathrm{part}$. In the  
color glass condensate approach \cite{Kharzeev:2000ph} this non-linear
rise is controlled by the centrality dependence of the saturation momentum
scale $Q_s$, with no free parameters to tune. It remains to be seen whether
the success of the CGC model in describing the centrality dependence of
$\dNdy$ \cite{Kharzeev:2001gp} survives the inclusion of entropy (or 
multiplicity) producing effects resulting from shear viscosity during the 
evolution from the initial CGC to the finally observed state. 

%%%%%%%%%%%%%%%%%%%%%%%%%%%%%%%%%%%%%%%%%%%%%%%%%%%%%%%%%%%%%%%%%%%%%%%%%%%%%
\section{Concluding remarks}
\label{sec7}
%%%%%%%%%%%%%%%%%%%%%%%%%%%%%%%%%%%%%%%%%%%%%%%%%%%%%%%%%%%%%%%%%%%%%%%%%%%%%

The main motivation for the work presented in this paper was provided
by the experimentally observed multiplicity scaling of the elliptic flow,
shown in Fig.~\ref{F7}a, and its deviation at low multiplicities from 
ideal fluid dynamical predictions. We saw that many of the observed 
features are qualitatively consistent with viscous hydrodynamic 
calculations as presented in this paper, and that the same calculations
also predict approximate multiplicity scaling for viscous entropy 
production. Our studies revealed, however, that even for ideal
fluid dynamics the multiplicity scaling of the elliptic flow is not 
perfect, with small scaling violations introduced by the freeze-out 
process which cuts the evolution of elliptic flow short. Even at RHIC 
energies, where the elliptic flow almost saturates before freeze-out, 
kinetic decoupling truncates the momentum anisotropy at values slightly
below their asymptotic saturation value, and the deviations depend on 
the size of the colliding nuclei and the deformation of the fireball
created in the collision through the time available for building 
elliptic flow before freeze-out.

Shear viscosity strongly suppresses the build-up of momentum anisotropy
and elliptic flow, especially for low multiplicity densities, i.e. at large
impact parameters, low collision energies or for small sizes of the colliding
nuclei. This changes the slope of the multiplicity scaling curve for
$v_2/\ecc$ but preserves, to good approximation, its general scaling
with $\frac{1}{S}\frac{dN_\mathrm{ch}}{dy}$. Violations of multiplicity 
scaling for $v_2/\ecc$ are somewhat larger for the viscous expansion 
than for the ideal fluid (especially with EOS~I), but remain small enough 
to be consistent, within statistical errors, with the experimental 
observation of approximate scaling. The slope of the approximate scaling 
curve and the spread around this curve caused by scaling violation 
increase with the value of the specific shear viscosity $\eta/s$ and 
can thus be used to constrain it.

Specifically, the observed scaling violations have the following features: 
At fixed multiplicity density $\frac{1}{S}\frac{dN_\mathrm{ch}}{dy}$,
viscous hydrodynamics predicts slightly larger elliptic flow $v_2/\ecc$
for larger collision systems or more central collisions than for smaller
nuclei colliding at similar energy or more peripheral collisions between
similar-size nuclei colliding at higher energy. Larger $v_2/\ecc$
values are associated with longer lifetimes of the corresponding fireballs 
before freeze-out and thus also with larger relative entropy production.
This correlates the scaling violations for $v_2/\ecc$ observed in 
Figs.~\ref{F5} and \ref{F7} with those for the relative entropy 
production $\Delta{\cal S}/{\cal S}_0$ seen in Fig.~\ref{F8}. The 
pattern of the predicted scaling violations shows qualitative agreement
with experiment, although higher quality data are required to
render this agreement statistically robust and quantitative.

For a fixed (i.e. temperature independent) ratio $\eta/s$, the slope of 
the multiplicity scaling curve for $v_2/\ecc$ does not agree with experiment 
-- the curves predicted by viscous hydrodynamics are too flat. The slope can
be increased by allowing $\eta/s$ to increase at lower temperatures:
For small multiplicity densities (very peripheral collisions or low 
collision energies), the data seem to require $\frac{\eta}{s}>3\times
\frac{1}{4\pi}$, whereas at large multiplicity densities they appear to 
constrain the specific shear viscosity to values of $\frac{\eta}{s}
\lesssim(1{-}2)\times\frac{1}{4\pi}$. While this is qualitatively
consistent with the idea that in high-multiplicity events the dynamics
is dominated by the QGP phase (whose viscosity would thus have to be small,
of order $1/4\pi$) whereas low-multiplicity events are predominantly
controlled by hadron gas dynamics (which is highly viscous 
\cite{Hirano:2005xf}), much additional work is needed to turn this 
observation into quantitative constraints for the function 
$\frac{\eta}{s}(T)$.

The present study also resolves questions that arose from several 
recent publications of viscous hydrodynamic calculations which seemed
to yield different results. We explored the effects of using different
implementations of Israel-Stewart second order theory for causal 
relativistic viscous hydrodynamics, by comparing the ``simplified 
Israel-Stewart equations'' previously used by us \cite{Song:2007fn}
with the ``full Israel-Stewart equations'' implemented by 
P. \& U. Romatschke \cite{Romatschke:2007mq}. For the ``simplified''
approach we found a strong sensitivity of physical observables on the
presently unknown kinetic relaxation time $\tau_\pi$ for the viscous 
shear pressure tensor $\pi^{mn}$, in contrast to a much weaker and 
basically negligible $\tau_\pi$-dependence in the ``full'' approach. 
For non-zero $\tau_\pi$ the ``simplified I-S equations'' allow for 
large excursions of $\pi^{mn}$ away from its Navier-Stokes limit
$\pi^{mn}=2\eta\sigma^{mn}$. These excursions are artificial and disappear
in the Navier-Stokes limit $\tau_\pi\to0$ which can, however, not be 
stably simulated numerically. They cause large viscous suppression 
effects for the elliptic flow and large amounts of extra entropy 
production (i.e. extra final hadron multiplicity). From our study we
conclude that the ``simplified I-S approach'' should be avoided, and
that a reliable extraction of $\eta/s$ from experimental data mandates 
the use of the ``full Israel-Stewart equations'' \cite{Romatschke:2007mq,%
Luzum:2008cw}. (It is, however, permissible to use the conformal fluid
approximation \cite{Romatschke:2007mq,Baier:2007ix} for the ``full I-S 
equations'' even if the fluid's EOS is not conformally invariant since
the differences were found to be negligible.)

In comparing our previous work \cite{Song:2007fn} with that of others
we also identified other factors that significantly influence the creation
of elliptic flow and thus help to account for the observed differences.
For a realistic equation of state that implements a quark-hadron transition
(here SM-EOS~Q and EOS~L), it turns out that a much more important effect
than using the correct version of Israel-Stewart theory is the size of
the colliding nuclei. At RHIC energies and for a realistic EOS, the 
viscous suppression effects for $v_2/\ecc$ in Cu+Cu collisions are 
almost twice as large as for the larger Au+Au collision system. Non-negligible
differences in the amount of viscous $v_2$ suppression arise also from 
details in the EOS, with a smooth crossover as implemented in EOS~L 
giving 25-30\% less suppression than a first-order transition as in 
SM-EOS~Q. Compared to system size effects and EOS uncertainties,
the differences between ``simplified'' and ``full'' I-S theory are 
relatively small, affecting the viscous $v_2$ suppression at the 10\%
level relative to each other. (The quoted percentages are for a fluid with
minimal viscosity $\eta/s=1/4\pi$ and may be larger for higher viscosity.)
The largest uncertainty, in any case, is contributed by our present lack
of knowledge of the initial source eccentricity which contributes a 
theoretical error band of up to 30\% on an absolute scale for $v_2$
\cite{Hirano:2005xf,Luzum:2008cw,Drescher:2006pi,Lappi:2006xc}.

We finally comment that the multiplicity dependence of viscous entropy 
production predicted by viscous hydrodynamics (see Fig.~\ref{F8}) will 
modify the centrality dependence of charged hadron production. This 
issue will be studied more quantitatively in a forthcoming paper. 

\acknowledgments
We thank K. Dusling, E. Frodermann, P. Huovinen, M. Lisa, P. Romatschke, 
and D. Teaney for fruitful discussions and useful comments on the 
manuscript. This work was supported by the U.S. Department of Energy 
under contract DE-FG02-01ER41190.

%%%%%%%%%%%%%%%%%%%%%%%%  References %%%%%%%%%%%%%%%%%%%%%%%%%%%%%%%%%%%%%%%%%

\end{document}